\documentclass[twocolumn,floatfix,prb]{revtex4-2}%
\usepackage{graphicx}
\usepackage{bmpsize}
\usepackage{amsmath}%
\usepackage{amsfonts}%
\usepackage{amssymb}
\usepackage{bm}
\usepackage{color}
\usepackage[usenames,dvipsnames]{xcolor}
\usepackage{ulem}
\usepackage{comment}
\usepackage{braket}
\usepackage{color}
\usepackage[top=30truemm,bottom=30truemm,left=25truemm,right=25truemm]
{geometry}

\def\0{{ {\bm 0} }}

\allowdisplaybreaks[4]

\renewcommand{\thefigure}{\arabic{figure}}
\renewcommand{\theequation}{\arabic{equation}}

\begin{document}
\title{
Surface State of Inter-orbital Pairing State in $\mathrm{Sr_2RuO_4}$ Superconductor}
\author{Satoshi Ando, Satoshi Ikegaya, Shun Tamura, Yukio Tanaka, Keiji Yada}
\date{\today }
\begin{abstract}
  We study the (001) surface state of a recently proposed $E_g$ symmetry inter-orbital-odd spin-triplet $s$-wave superconducting (SC) state in $\mathrm{Sr_2RuO_4}$ (SRO).
  We confirm that this pair potential is transformed into a chiral $d$-wave pair potential and a pseudo-Zeeman field in the band basis for a low-energy range.
  Because of the chiral $d$-wave pair potential, the surface states appear near zero energy in the momentum range enclosed by the nodal lines of the chiral $d$-wave pair potential for each band at the (001) surface. 
  Nevertheless, the pseudo-Zeeman field gives band splitting of the surface states, and its splitting energy is much smaller than the SC energy gap.
  The local density of states (LDOS) at the (001) surface of the SC state has a pronounced peak structure at zero energy because of the surface states near zero energy when the order of the resolution is lower than the splitting energy. 
  This peak structure is robust against perturbations, such as an orbital Rashba coupling or an $E_u$ SC pair potential at the surface.
\end{abstract}
\address{
Department of Applied Physics,
Nagoya University,
Nagoya 464-8603,
Japan
}
\sloppy
\maketitle
\section{Introduction}
 The superconducting (SC) symmetry in $\mathrm{Sr_2RuO_4}$ (SRO) has been a central issue in condensed-matter physics since its discovery \cite{SRO-Y-Maeno}. 
 Spin-triplet chiral $p$-wave [$k_x+\mathrm{i}k_y$-wave] was long considered the leading candidate for SC symmetry because of the observations of the constant spin-susceptibility (NMR \cite{K-Ishida} and neutron scattering \cite{J-A-Duffy}) and time-reversal symmetry breaking (TRSB) ($\mu$SR \cite{muSR} and Kerr effect \cite{Kerr}).
 In addition, numerous theoretical studies have supported the realization of the $k_x+\mathrm{i}k_y$-wave state \cite{T-Nomura,M-Sato2,T-Takimoto,K-Kuroki,T-Nomura2,T-Nomura3,Y-Yanase,T-Nomura4,T-Nomura5,S-Raghu,M-Tsuchiizu,L-D-Zhang,W-S-Wang,Z-Wang}. 
 Recently, however, Pustogow et al. \cite{A-Pustogow} pointed out that there had been a heating issue in the previously reported NMR experiment. 
When this problem was solved, spin susceptibility under an in-plane magnetic field was suppressed at temperatures below $T_{c}$ in both NMR and $\mu$SR measurements \cite{A-Pustogow,K-Ishida2,A-Chronister,A-N-Petsch}.
 These recent results appear to be inconsistent with the $k_x+\mathrm{i}k_y$-wave state.
 However, ultrasound and thermodynamics experiments suggest multiple degenerate order parameters \cite{S-Benhabib,D-F-Agterberg, S-Ghosh}, and several theoretical studies focused on the two-component order parameters with TRSB in SRO have been reported ~\cite{S-Ikegaya,A-T-Romer,J-Clepkens2, A-T-Romer2,S-A-Kivelson, R-Willa,J-Clepkens, A-C-Yuan,A-Ramires,H-G-Suh, S-Kaser}.
 In general, TRSB SC states are composed of two different components that belong to two different irreducible representations (irreps) or to the same irrep. 
  In cases where two different irreps are mixed, the possibility of certain combinations has been proposed, such as $s'+\mathrm{i}d_{x^2-y^2}$-wave \cite{A-T-Romer}, $d_{x^2-y^2}+\mathrm{i}g_{xy(x^2-y^2)}$-wave \cite{S-A-Kivelson, R-Willa, J-Clepkens, A-C-Yuan},  and $s+\mathrm{i}d_{xy}$-wave \cite{J-Clepkens2, A-T-Romer2}. 
However, in the same irreps case, the allowed pair symmetry has been reported to be only $E_g$ irrep 
by Ref. \cite{A-Pustogow,K-Ishida2,A-Chronister,A-N-Petsch} in $D_{4h}$ \cite{Siglist}. 
The most simple basis function of $E_g$ symmetry is the $k_z(k_x+\mathrm{i}k_y)$-wave (i.e., the chiral $d$-wave).
However, this simple chiral $d$-wave pairing forms Cooper pairs between the electrons in different layers of an SRO crystal, which has a nearly two-dimensional electronic structure; 
therefore, the formation of such Cooper pairs has been considered difficult.
Nevertheless, when the orbital degree of freedom is considered, onsite $E_g$ SC pairing states, which can resolve the aforementioned problems, are possible. 
Actually, an inter-orbital-odd spin-triplet  $s$-wave pairing state in the $E_g$ irrep has recently attracted attention~\cite{A-Ramires,H-G-Suh, S-Kaser}. 
This pair has an energy-gap structure similar to that of a chiral $d$-wave pairing but differs in that it has a Bogoliubov Fermi surface. 
This state is one of the most promising candidates because it can explain the $\mu$SR experiments under both hydrostatic pressure \cite{hydrostatic} and in-plane uniaxial strain \cite{uniaxial}.

In this paper, we calculate 
the dispersion of the surface state and the local density of states (LDOS) at the (001) surface (hereafter referred to as the top surface) of the $E_g$ inter-orbital SC state by the recursive Green's function formula \cite{A-Umerski}.
The surface states appear near zero energy with splits by a much smaller energy scale than that of the SC gap.
The physical origin of this surface state can be understood by an effective low-energy Hamiltonian characterized by an effective chiral $d$-wave pair potential and a pseudo-Zeeman field. 
As a result, the LDOS has a pronounced zero-energy peak when the order of the resolution is lower than the splitting energy.
This peak structure is robust against perturbations at the top surface, such as an orbital Rashba coupling or a chiral $p$-wave SC pair potential.

\section{Analytical Description}
\subsection{Model Hamiltonian}
Let us start with a model Hamiltonian for an inter-orbital superconducting state of SRO, as originally proposed in Ref.~\cite{H-G-Suh}.
We focus on the $t_{2g}$ orbitals of the Ru ions (i.e., the $d_{yz}$-, $d_{zx}$-, and $d_{xy}$-orbitals), which dominate the bands near the Fermi level.
The corresponding Bogoliubov--de Gennes (BdG) Hamiltonian is given by
\begin{align}
H=\frac{1}{2}\sum_{\boldsymbol{k}}
\Psi^{\dagger}_{\boldsymbol{k}}\check{H}({\boldsymbol{k}})\Psi_{\boldsymbol{k}}
\end{align}
with
\begin{align} \label{eq:BdG}
\check{H}({\boldsymbol{k}})=\left(\begin{array}{cc}
\hat{H}_N({\boldsymbol{k}}) & \hat{\Delta}({\boldsymbol{k}}) \\ \hat{\Delta}^{\dagger}({\boldsymbol{k}}) & -\hat{H}_N^{\mathrm{T}}(-{\boldsymbol{k}}) \end{array} \right),
\end{align}
and
\begin{align}
\begin{split}
&\Psi^{\dagger}_{\boldsymbol{k}}=(\boldsymbol{C}^{\dagger}_{\boldsymbol{k}}, \boldsymbol{C}^{\mathrm{T}}_{-\boldsymbol{k}}),\\
&\boldsymbol{C}^{\dagger}_{\boldsymbol{k}}
=(c^{\dagger}_{\boldsymbol{k},\uparrow,yz},c^{\dagger}_{\boldsymbol{k},\uparrow,zx},c^{\dagger}_{\boldsymbol{k},\uparrow,xy}
c^{\dagger}_{\boldsymbol{k},\downarrow,yz},c^{\dagger}_{\boldsymbol{k},\downarrow,zx},c^{\dagger}_{\boldsymbol{k},\downarrow,xy}),
\end{split}
\end{align}
where $c^{\dagger}_{\boldsymbol{k},\sigma,\chi}$ creates an electron with momentum $\boldsymbol{k}$ and spin $\sigma$ in orbital $\chi$.
The normal-state Hamiltonian is described by
\begin{align}
\begin{split}
&\hat{H}_N({\boldsymbol{k}})=\left(\begin{array}{cc}
\bar{\xi}+\bar{\lambda}_0+\bar{\lambda}_3 & \bar{\lambda}_1+\mathrm{i} \bar{\lambda}_2 \\
\bar{\lambda}_1-\mathrm{i} \bar{\lambda}_2 & \bar{\xi}+\bar{\lambda}_0-\bar{\lambda}_3 \end{array} \right),\\
&\bar{\xi}=\left(\begin{array}{ccc} \xi_{yz} & 0 & 0  \\  0 & \xi_{zx} & 0  \\  0 & 0& \xi_{xy} \end{array} \right),\quad
\bar{\lambda}_0 =\left(\begin{array}{ccc} 0 & h_{10} & h_{20}  
  \\  h_{10} & 0 & h_{30}  \\  h_{20} &  h_{30} & 0 \end{array} \right),\\
&\bar{\lambda}_{j=1,2,3}=\left(\begin{array}{ccc} 0 & - \mathrm{i} h_{4j} & - \mathrm{i} h_{5j}  \\  \mathrm{i} h_{4j} & 0 & - \mathrm{i} h_{6j} \\  \mathrm{i} h_{5j} & \mathrm{i} h_{6j} & 0 \end{array} \right),
\label{eq:ham_normal}
\end{split}
\end{align}
where $\bar{\xi}+\bar{\lambda}_0$ contains the spin-independent hopping integrals and the chemical potentials and $\bar{\lambda}_{j=1,2,3}$ describe the inter-orbital spin--orbit couplings.
The explicit forms of the matrix elements $h_{ij}$ and the corresponding band parameters are summarized in Appendix~\ref{ap:model}.
The proposed inter-orbital pair potential associated with the $E_g$ irrep is denoted by
\begin{align}
\begin{split}
\label{eq:pairpotential}
&\hat{\Delta}=\left(\begin{array}{cc} 0 & \bar{\Delta} \\ \bar{\Delta} & 0 \end{array} \right),\quad
\bar{\Delta} = \Delta \left(\bar{L}_x + \mathrm{i} \bar{L}_y\right),\\
&\bar{L}_x=\left(\begin{array}{ccc} 0 & 0 & 0  \\  0 & 0 & \mathrm{i}  \\  0 &  -\mathrm{i} & 0 \end{array} \right), \quad
\bar{L}_y=\left(\begin{array}{ccc} 0 & 0 & -\mathrm{i} \\  0 & 0 & 0  \\  \mathrm{i} &  0 & 0 \end{array} \right),
\end{split}
\end{align}
where $\bar{L}_x$ and $\bar{L}_y$ are the orbital angular momentum operators in the orbital space and $\Delta$ ($\geq 0$) denotes the magnitude of the pair potential.
We assume that $\Delta$ is constant with respect to $\boldsymbol{k}$.
The present pair potential describes spin-triplet superconductivity, where the $d$-vector is directed along the $c$-axis of SRO.
Notably, this inter-orbital superconducting state is stabilized by the spin--orbit couplings of $h_{53}$ and $h_{63}$~\cite{H-G-Suh}.

\subsection{Approximate low-energy Hamiltonian}
\begin{figure*}[htbp]
\includegraphics[width=1\linewidth]{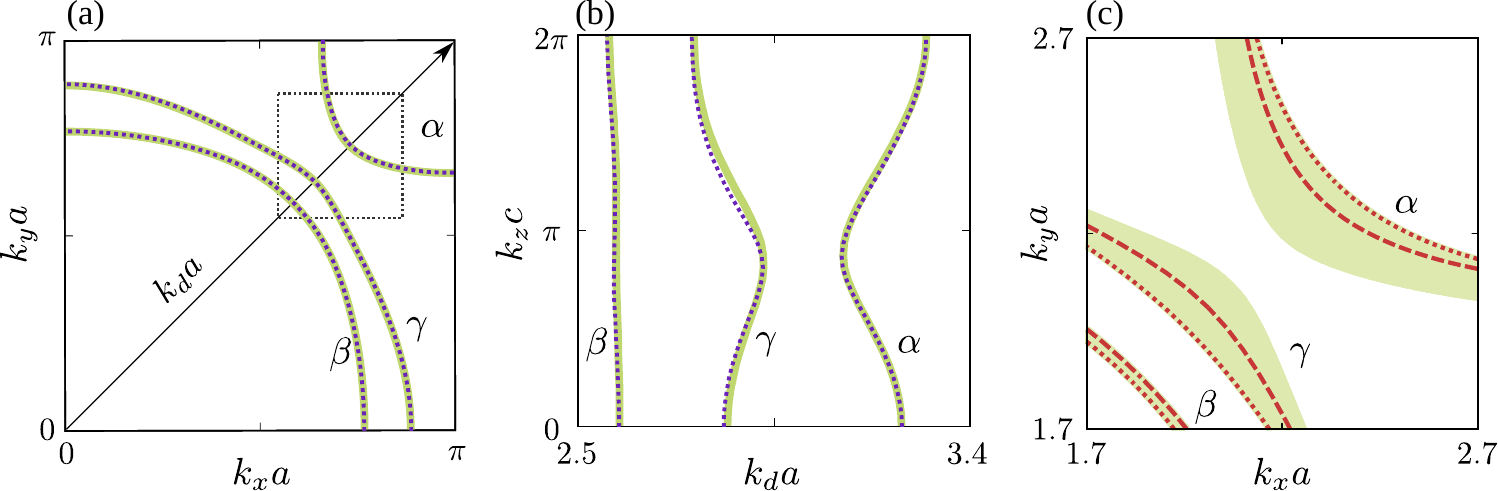}
\caption{(Color online)
(a) The Fermi surface at $k_zc=0$, where the arrow shows the $k_da$ axis, which is the diagonal line on the $k_xk_y$ plane. 
(b) The $k_z$ dependence of the Fermi surface on the $k_da$ line shown in figure~(a).
Solid lines are in the approximated band basis,
and dotted lines are in the numerical band basis. 
(c) The numerical Fermi surface projected into the surface Brillouin zone in the range enclosed by the square shown in figure~(a).
Dotted and dashed lines show the numerical Fermi surface at $k_zc=2\pi$ and $k_zc=0$, respectively.} 
\label{fig:FS}
\end{figure*}
In this subsection, we derive an approximate Hamiltonian that enables us to grasp the essential properties of the present model.
For this purpose, although we lose quantitative accuracy, we treat the inter-orbital hybridizations of $\bar{\lambda}_{i=0,1,2,3}$ as the perturbation.
On the basis of the second-order perturbation theory, we can deform the Hamiltonian in an approximate band basis as
\begin{align}
&\check{u}^{\dagger}\check{H}({\boldsymbol{k}})\check{u}
=\left(\begin{array}{ccc} \breve{\mathcal{H}}_{\alpha} & \breve{\mathcal{V}}_{\alpha \beta} & \breve{\mathcal{V}}_{\alpha \gamma}  \\ 
\breve{\mathcal{V}}_{\beta \alpha } & \breve{\mathcal{H}}_{\beta} & \breve{\mathcal{V}}_{\beta \gamma}  \\ 
\breve{\mathcal{V}}_{\gamma \alpha} &  \breve{\mathcal{V}}_{\gamma \beta} & \breve{\mathcal{H}}_{\gamma} \end{array} \right) + O(\lambda^3),
\label{eq:band_ham}
\end{align}
  with
  \begin{align} \label{eq:chiral_d_part}
  \begin{split}
  &\breve{\mathcal{H}}_{\nu}=\left(\begin{array}{cc}
  \tilde{\varepsilon}_{\nu} & \tilde{\psi}_{\nu}\\
  \tilde{\psi}^{\dagger}_{\nu} &-\tilde{\varepsilon}_{\nu} \end{array}\right),\\
  &\tilde{\varepsilon}_{\nu}=\varepsilon_{\nu} \tilde{\sigma}_0,\quad
  \tilde{\psi}_{\nu}=\psi_{\nu}(\mathrm{i}\tilde{\sigma}_2),
  \end{split}
  \end{align}  
and
\begin{align}
  \begin{split}
  &\breve{\mathcal{V}}_{\nu\nu^{\prime}}=\left(\begin{array}{cc}
  0 & \tilde{\mathcal{D}}_{\nu\nu^{\prime}}\\
  -\tilde{\mathcal{D}}^{\ast}_{\nu\nu^{\prime}} & 0\end{array}\right),\\
  &\tilde{\mathcal{D}}_{\nu\nu^{\prime}}=\left(\psi_{\nu\nu^{\prime}}+\boldsymbol{d}_{\nu\nu^{\prime}}\cdot \tilde{\boldsymbol{\sigma}}\right)(\mathrm{i}\tilde{\sigma}_2),\\
  &\boldsymbol{d}_{\nu\nu^{\prime}}=(d_{1,\nu\nu^{\prime}},d_{2,\nu\nu^{\prime}},d_{3,\nu\nu^{\prime}}),
  \end{split}
  \end{align}
  for $\nu$, $\nu^{\prime}=\alpha$, $\beta$, and $\gamma$, where $\tilde{\boldsymbol{\sigma}}=(\tilde{\sigma}_1,\tilde{\sigma}_2,\tilde{\sigma}_3)$ and $\tilde{\sigma}_0$ represent the Pauli matrices and the unit matrix in pseudo-spin space, respectively, and $O({\lambda}^n)$ represents the Landau symbol with respect to the $n$-th order of the matrix elements in $\bar{\lambda}_{i}$.
  The explicit forms of the matrix elements and the unitary operator $\check{u}$ are given in Appendix~\ref{ap:low-energy}.
  We construct the unitary matrix $\check{u}$ to diagonalize the normal-state Hamiltonian $\hat{H}_N$ within the second-order of $\bar{\lambda}_{i}$.
  In addition, the diagonal components of $\varepsilon_{\alpha}$, $\varepsilon_{\beta}$, and $\varepsilon_{\gamma}$ give the kinetic energies of three different bands, which constitute three separated Fermi surfaces.
  Thus, the unitary transformation in Eq.~(\ref{eq:band_ham}) changes the basis of the Hamiltonian from the original orbital basis to the approximate band basis.
  Then,  $\psi_{\nu,\nu'}\mathrm{i}\tilde{\sigma}_2$ and $\bm{d}_{\nu,\nu'}\cdot\tilde{\bm{\sigma}}\mathrm{i}\tilde{\sigma}_2$ ($\bm{d}_{\nu,\nu}=\bm{0}$) are the spin-singlet and -triplet pair potentials, respectively. 
  Parameters  $d_{1,\nu\nu^{\prime}},d_{2,\nu\nu^{\prime}},$ and $d_{3,\nu\nu^{\prime}}$ are the $x$-, $y$-, and $z$-components of the $d$-vector, respectively. 
These pair potentials have momentum dependence even though the original pair potential does not. 
  To check the validity of this approximation, we show the Fermi surfaces in the approximate band basis and the numerical band basis, which are obtained by the numerical diagonalization of the normal-state Hamiltonian in Figs.~\ref{fig:FS}(a) and (b).
  In Fig.~\ref{fig:FS}, the lattice constant of the conventional unit cell in the in-plane directions and the $z$-direction are represented by $a$ and $c$, respectively, and $c$ is twice the distance between nearest-neighbor layers.
  The approximate Fermi surfaces (dotted lines) and the numerical Fermi surfaces (solid lines) nicely correspond. 
  Figure~\ref{fig:FS}(c) shows the numerical Fermi surfaces projected onto the surface Brillouin zone (BZ).
  The dashed and dotted lines indicate the Fermi surfaces at $k_zc=0$ and $k_zc = 2\pi$, respectively.
  As we will show later, the property of the projected Fermi surfaces is related to the emergence of the low-energy surface states at the top surface.
  
 We here proceed with a further approximation to construct a low-energy effective Hamiltonian for each band.
  According to the argument in Refs.~\cite{P-M-R-Brydon,D-F-Agterberg-Bogoliubov,H-G-Suh,J-W-F-Venderbos}, the low-energy excitation in the vicinity of the Fermi surface of the $\nu$-band can be evaluated by
\begin{align}
    \begin{split}
        \breve{\mathcal{H}}^{\mathrm{eff}}_{\nu}=\breve{\mathcal{H}}_{\nu}
        +\sum_{\nu'\neq\nu}\frac{\breve{\mathcal{V}}^{\dag}_{\nu'\nu} \breve{\tau_z} 
        \breve{\mathcal{V}}_{\nu'\nu}}{\varepsilon_\nu-\varepsilon_{\nu'}}. \\
    \end{split}
    \label{eq:eff_ham0}
\end{align}
  where $\breve{\tau}_z=\mathrm{diag}[1,1,-1,-1]$.
  A detailed derivation of Eq.~(\ref{eq:eff_ham0}) is presented in Appendix~\ref{ap:low-energy}.
  Eventually, we obtain the low-energy effective Hamiltonian for the $\nu$-band within the second-order perturbation of $\bar{\lambda}_{i}$:
  \begin{align}
  \begin{split}
  &\breve{\mathcal{H}}_{\nu}^{\mathrm{eff}}=\left[\begin{array}{cc}
  \tilde{h}_{\nu}(\boldsymbol{k}) & \tilde{\psi}_{\nu}(\boldsymbol{k})\\
  \tilde{\psi}_{\nu}^{\dagger}(\boldsymbol{k}) &-\tilde{h}_{\nu}^{\mathrm{T}}(-\boldsymbol{k}) \end{array}\right],\\
  &\tilde{h}_{\nu} = \left(\varepsilon_{\nu} - \gamma_{\nu} \right)\tilde{\sigma}_0 + \boldsymbol{m}_{\nu}\cdot \tilde{\boldsymbol{\sigma}},\\
  \end{split}
  \end{align}
  with
  \begin{align}
  \label{eq:sift}
 \gamma_{\nu}=\sum_{\nu^{\prime}\neq \nu}  \frac{|\psi_{\nu^{\prime}\nu}|^2+|\boldsymbol{d}_{\nu^{\prime}\nu}|^2}
  {\varepsilon_{\nu}-\varepsilon_{\nu^{\prime}}},
  \end{align}
  \begin{align}
  \label{eq:zeeman}
\boldsymbol{m}_{\nu}=\sum_{\nu^{\prime} \neq \nu}
  \frac{2\mathrm{Re}[\psi_{\nu^{\prime}\nu}\boldsymbol{d}^{\ast}_{\nu^{\prime}\nu}]-\mathrm{i}\boldsymbol{d}_{\nu^{\prime}\nu}\times \boldsymbol{d}^{\ast}_{\nu^{\prime}\nu}}
{\varepsilon_{\nu}\varepsilon_{\nu^{\prime}}},
  \end{align}
  where $\gamma_{\nu}$ describes the modulation in the kinetic energy and $\boldsymbol{m}_{\nu}$ represents the pseudo-Zeeman potential.
  
  We now describe the essential properties of the present model, which is clarified using the effective low-energy Hamiltonian $\breve{\mathcal{H}}_{\nu}^{\mathrm{eff}}$.
  Remarkably, the effective pair potential $\psi_{\nu}$ acting on each band can be decomposed as
  \begin{align}
  \label{psi_1}
  \psi_{\nu}=X_{\nu}(\boldsymbol{k})+\mathrm{i}Y_{\nu}(\boldsymbol{k}),
  \end{align}
  where the real functions of $X_{\nu}(\boldsymbol{k})$ and $Y_{\nu}(\boldsymbol{k})$ obey
  \begin{align}
  X_{\nu}(-k_x,&k_y,k_z) = X_{\nu}(k_x,k_y,-k_z)=-X_{\nu}(\boldsymbol{k}),\nonumber\\
  &X_{\nu}(k_x,-k_y,k_z)=X_{\nu}(\boldsymbol{k}),
  \end{align}
  and
  \begin{align}
  \label{psi_2}
  Y_{\nu}(k_x,-&k_y,k_z) = Y_{\nu}(k_x,k_y,-k_z)=-Y_{\nu}(\boldsymbol{k}),\nonumber\\
  &Y_{\nu}(-k_x,k_y,k_z)=Y_{\nu}(\boldsymbol{k}),
  \end{align}
  respectively.
  This implies that the effective pair potential $\psi_{\nu}$ has chiral $d$-wave pairing symmetry similar to the $k_z(k_x+\mathrm{i}k_y)$-wave pair potential.
  The pair potential vanishes at $k_zc=0$ and $k_z c=2\pi$.
  Thus, in the absence of the pseudo-Zeeman potential (i.e., $\boldsymbol{m}_{\nu}=0$), $\breve{\mathcal{H}}_{\nu}$ exhibits the line nodes at $k_z=0$ and $k_z c=2\pi$, where the location of the line nodes corresponds to the dashed and dotted lines in Fig.~\ref{fig:FS}(c).
  In addition, the pair potential has mirror-odd nature with respect to $k_z$ (i.e., $\psi_{\nu}(k_x,k_y,k_z)=-\psi_{\nu}(k_x,k_y,-k_z)$), which enables us to expect the formation of low-energy surface states at the top surface~\cite{S-Kobayashi,hu_94,asano_04,M-Sato,kashiwaya_00,fu_18,Y-Tanaka4,G-Zwicknagl,J-Hara}.
  Moreover, it has been already shown that, for a $k_z(k_x+\mathrm{i}k_y)$-wave superconductor, we obtain topologically-protected flat-band zero-energy surface states at the top surface~\cite{S-Kobayashi,tamura_17,suzuki_20}.
  More specifically, the relevant topological invariant predicts that the flat-band zero-energy surface states appear at momenta enclosed by the nodal lines in the surface BZ (see also the detailed discussion in Appendix~\ref{ap:chiral_d}).
  Thus, on the basis of the analogy between the $k_z(k_x+\mathrm{i}k_y)$-wave superconductor and the present superconductor characterized by the effective chiral $d$-wave pair potential, we expect that, if the pseudo-Zeeman potential is absent, flat-band zero-energy surface states will appear in the momentum range enclosed by the dashed and dotted lines in Fig.~\ref{fig:FS}(c).
  Nevertheless, in the present model, the emergence of the pseudo-Zeeman potential is inevitable.
  The energy eigenvalue of $\breve{\mathcal{H}}^{\mathrm{eff}}_{\nu}$ is given by
  \begin{align}
  \begin{split}
  &E_{\pm,s}= \pm E_{\mathrm{cd}} + s|\boldsymbol{m}_{\nu}|,\\
  &E_{\mathrm{cd}}=\sqrt{(\varepsilon_{\nu} + \gamma_{\nu})^2+|\psi_{\nu}|^2},
  \end{split}
  \end{align}
  for $s=\pm$.
  The pseudo-Zeeman potential clearly shifts the bands of $\pm E_{\mathrm{cd}}$, which originally exhibit the line nodes at momenta satisfying $\varepsilon_{\nu} + \gamma_{\nu}=0$ and $|\psi_{\nu}|=0$.
  In particular, the line nodes in $E_{+,-}= E_{\mathrm{cd}}-|\boldsymbol{m}_{\nu}|$ and $E_{-,+}= - E_{\mathrm{cd}}+|\boldsymbol{m}_{\nu}|$ are inflated to the Bogoliubov--Fermi surfaces by the pseudo-Zeeman field~\cite{D-F-Agterberg-Bogoliubov,P-M-R-Brydon}.
  Moreover, we infer that exact flat-band zero-energy surface states can no longer exist because of the energy splitting from the pseudo-Zeeman field.
  Even so, the energy splitting in the surface states would be substantially smaller than $\Delta$ because the pseudo-Zeeman potential $|\boldsymbol{m}_{\nu}|$ is proportional to $\Delta^2$.
  
   Summarizing the previous discussion, we can expect the present model to display nearly zero-energy surface states at the top surface in the momentum range enclosed by the dashed and dotted lines in Fig.~\ref{fig:FS}(c), where the energy splitting of the surface states is due to the pseudo-Zeeman potential $\boldsymbol{m}_{\nu} \propto \Delta^2$.
  In the next section, we will confirm this statement by examining the surface energy dispersion and the surface LDOS numerically.

\section{Detailed Properties of \\Surface States}
\subsection{Recursive Green's function techniques}
In this section, we consider the open boundary condition in the $z$-direction and the periodic boundary condition in the $x$- and $y$-directions to calculate the surface Green's function for the semi-infinite system.
In addition, we consider flat surfaces and do not consider surface reconstructions.
Then, the momentum parallel to the surface $\bm{k}_\parallel\equiv (k_x, k_y)$  becomes a good  quantum number, and the problem is reduced to a one-dimensional problem along the $z$-direction at each momentum $\bm{k}_\parallel$.
The BdG Hamiltonian $\check{H}(\bm{k})$ in Eq.~(\ref{eq:BdG}) includes inter-layer hopping up to the next-nearest layer.
Thus, the Hamiltonian of a system with $n$ layers stacked in the $z$-direction can be written as
\begin{align}
\label{eq:n_layer}
H_n(\bm{k}_\parallel)
=&
\sum_{j=1}^{n}\sum_{\alpha,\beta} C_{j,\alpha,\bm{k}_\parallel}^{\dagger} \{h_0(\bm{k}_\parallel)\}_{\alpha,\beta}
C_{j,\beta,\bm{k}_\parallel}
\nonumber
\\
+&
\sum_{j=1}^{n-1}\sum_{\alpha,\beta} [C_{j,\alpha,\bm{k}_\parallel}^\dagger  \{t_1(\bm{k}_\parallel)\}_{\alpha,\beta}
C_{j+1,\beta,\bm{k}_\parallel} +\mathrm{H.c.}]
\nonumber\\
+&
\sum_{j=1}^{n-2}\sum_{\alpha,\beta} [C_{j,\alpha,\bm{k}_\parallel}^\dagger  \{t_2(\bm{k}_\parallel)\}_{\alpha,\beta} C_{j+2,\beta,\bm{k}_\parallel} +\mathrm{H.c.}],
\end{align}
where $\alpha$ and $\beta$ denote the internal degrees of freedom: spin, orbital, and particle--hole; $C_{j,\alpha,\bm{k}_\parallel}^\dagger$ and  $C_{j,\alpha,\bm{k}_\parallel}$ are the creation and annihilation operators at the $j$-th site in the $z$-direction, respectively, and $h_0(\bm{k}_\parallel)$ and $t_{1(2)}(\bm{k}_\parallel)$ are the intra-layer element and the inter-layer hopping between the (next) nearest-neighbor layers of the Hamiltonian, respectively.

To obtain the surface Green's function of the top layer, we use the recursive Green's function technique.
By formulating the recursion relation using the M\"{o}bius transformation, we can calculate the surface Green's function in a semi-finite system \cite{A-Umerski}.
This technique is developed for the system with up to nearest-neighbor hopping, whereas the Hamiltonian in Eq.~(\ref{eq:n_layer}) includes next-nearest-neighbor hopping.
Thus, 
we treat the $n$-layers system as the $n'(=n/2)$-layers system by treating two adjacent original layers as a single layer.
The Green's function of the $2n'$-layers system is given by $G^{n’}(z,\bm{k}_\parallel)={[z I-H_{2n’}(\bm{k}_\parallel)]}^{-1}$.
 Then, the $(n’,n’)$ and $(n’-1,n’-1)$ components of the Green’s functions $G_{n’}^{n’}$ and $G_{n’-1}^{n’}$ satisfy the next equation: 
\begin{align} 
&G_{n'}^{n'}(z,\bm{k}_\parallel)
\nonumber\\
&={\left[zI-h(\bm{k}_\parallel)-t^\dagger(\bm{k}_\parallel) G_{n'-1}^{n'}(z,\bm{k}_\parallel)t(\bm{k}_\parallel)\right]}^{-1},
\label{eq:recursive}
\end{align}
where     
$z=\varepsilon+\mathrm{i}\delta$ and $I$ is the unit matrix.
Parameter $\varepsilon$ is a real frequency, and $\delta$ is an infinitesimal imaginary part.
Here, $h(\bm{k}_\parallel)$ and $t(\bm{k}_\parallel)$ are given by  
\begin{align}
&{h} =  
\left(
    \begin{array}{cc}
      h_0              &
      t_1                   \\
      t_1^{\dag}                                          
                            &
      h_0                      
    \end{array}
  \right)
\end{align}
and
\begin{align}
\quad t =  
\left(
    \begin{array}{cc}
      t_2              &
      0                   \\
      t_1                                          
           &
      t_2                      
    \end{array}
  \right),
\end{align}
respectively.
By taking the limit $n'\rightarrow\infty$ with the M\"obius transformation, we obtain the top-surface Green's function for the semi-infinite system \cite{A-Umerski}: $\dot{g}^s=\lim_{{n'}\rightarrow\infty}G_{n'}^{n'}$, where $\dot{g}^s$ is a $24\times24$ matrix with the orbital, spin, particle--hole, and layer degrees of freedom. 
The details are given in Appendix~\ref{ap:recursive}.
The energy dispersion of the surface states within the bulk energy gap can be determined by examining the poles of the Green's function.
Thus, we can obtain the dispersion of the surface state from the pole of $\dot{g}^s$.
In the same manner, we can obtain the bottom-surface Green's function: $\dot{g}^{s,\mathrm{btm}}=\lim_{{n'}\rightarrow\infty}G_1^{n'}$
from
$G_1^{n'}
={\left[zI-h-t G_{2}^{n'}t^\dagger\right]}^{-1}$. 
From $\dot{g}^{s}$ and $\dot{g}^{s,\mathrm{btm}}$, we can obtain the Green's function in the bulk:
\begin{align}
\dot{g}={(zI-h-t^\dagger \dot{g}^s t-t \dot{g}^{s,\mathrm{btm}}t^\dagger)}^{-1}.
\end{align}

The LDOS of the top surface $N^s(\varepsilon)$ and the DOS of the bulk $N(\varepsilon)$ are given by
\begin{align} 
N^s(\varepsilon)=-1/N\sum_{\bm{k}_\parallel, i}
1/\pi[{\mathrm{Im}} \dot{g}^s_{ii}(\bm{k}_\parallel ,\varepsilon)],
\end{align}
\begin{align} 
  N(\varepsilon)=-1/N\sum_{\bm{k}_\parallel, i}
  1/\pi[{\mathrm{Im}} \dot{g}_{ii}(\bm{k}_\parallel ,\varepsilon)],
\end{align}
respectively.
Here, we take the summation of $i$ within the electron part in the one layer. 
Parameter $N$ is the number of sites in the $xy$-plane.

\subsection{Surface energy spectrum}
\begin{figure}[!htb]
\includegraphics[width=1.0\linewidth]{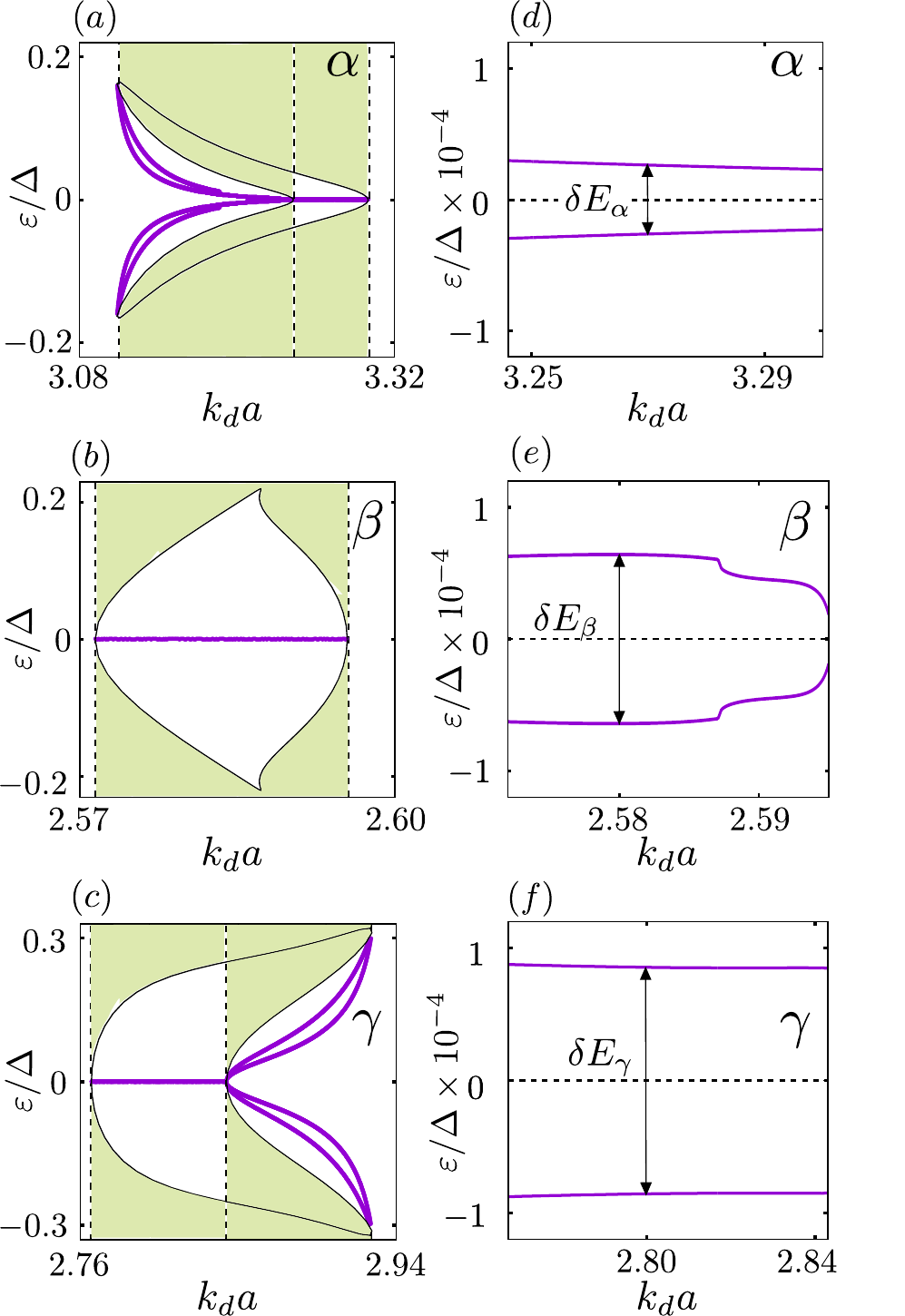}
\caption{(Color online)
The surface states in the (a) $\alpha$, (b) $\beta$, and (c) $\gamma$ bands.
Solid lines are the surface state, and shaded areas are the continuous energy levels in the bulk.
The surface states between the dotted and dashed lines are shown in Fig.~\ref{fig:FS}(c) in a tiny energy range for (d) $\alpha$, (e) $\beta$, and (f) $\gamma$ bands.
$\delta E_{\alpha,\beta,\gamma}$ are the width of the band splitting at $k_{d} a=3.27,2.58,2.80$.
}
\label{fig:trgsw}
\end{figure}
\begin{figure}[!htb]
  \includegraphics[width=1\linewidth]{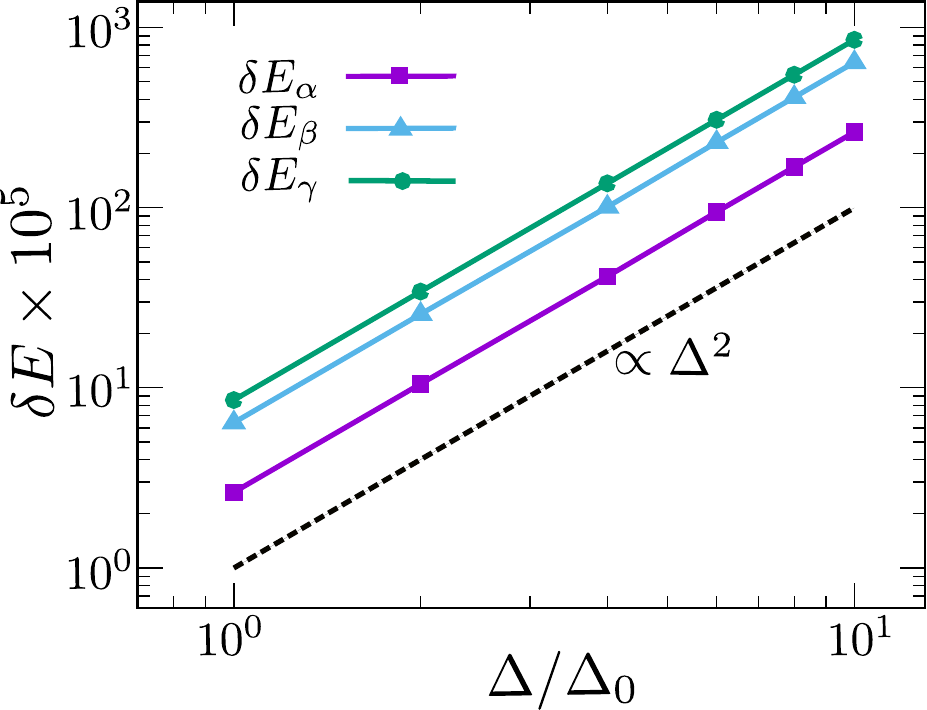}
  \caption{(Color online)
  Common logarithm plot of the $\Delta$ dependence of $\delta E_{\alpha,\beta,\gamma}$ shown in Figs. \ref{fig:trgsw}(d)--(f).
  The dotted line is proportional to $\Delta^2$. 
  $\Delta_0=|t_z^{(xy,xy)}\times10^{-4}|$.}
  \label{fig:split}
\end{figure}
\begin{figure*}[htbp]
\includegraphics[width=1\linewidth]{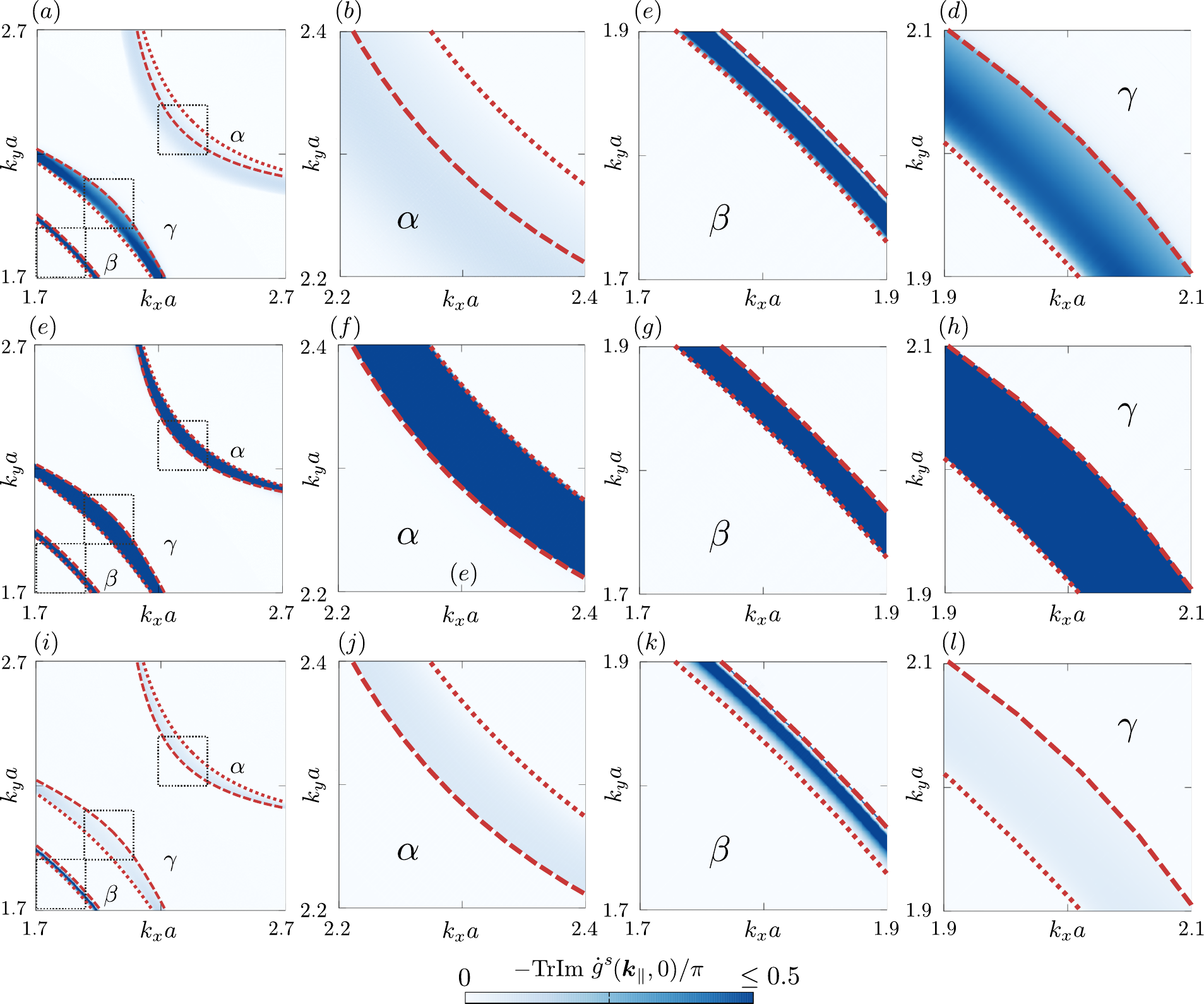}
\caption{(Color online)
The momentum-resolved LDOS at $\varepsilon=0$ in the momentum range shown in Fig. \ref{fig:FS}(c) for (a) $\delta=\Delta\times10^{-2}$, (e) $\delta=\Delta\times10^{-5}$, and (i) $\delta=\Delta\times10^{-7}$. (b)--(d), (f)--(g), and (j)--(l) are enlarged maps for (a), (e), and (i) for each band in the range surrounded by the squares, respectively.
Dotted and dashed lines show the Fermi surface at $k_zc=2\pi$ and $k_zc=0$, respectively.
}
\label{fig:trgsk}
\end{figure*}
\begin{figure}[!htb]
    \includegraphics[width=1\linewidth]{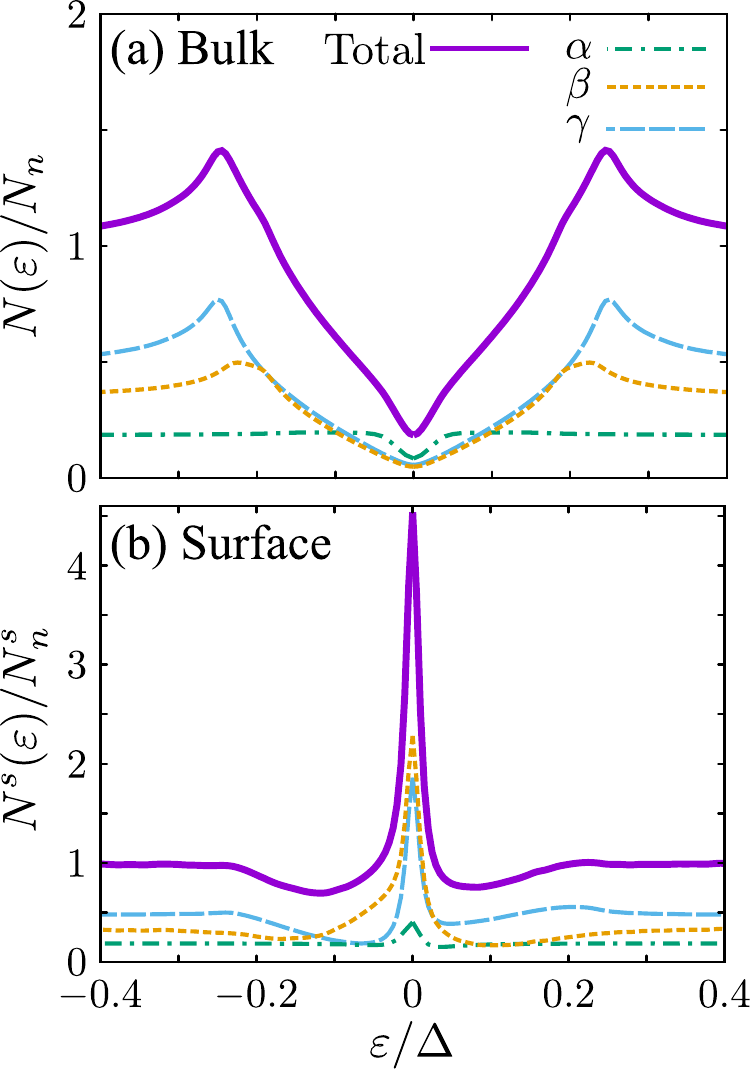}
    \caption{(Color online)
    (a) The DOS of the bulk and each band's contribution to the DOS.
    (b) The LDOS at the top surface and each band's contribution to the LDOS. }
    \label{fig:DOS}
\end{figure}
First, we consider the energy dispersion on the diagonal direction of the surface BZ.
We choose $\Delta$ to be the same order of $T_c$ in SRO.
$\Delta =|t_z^{(xy,xy)}\times10^{-4}|$, where $t_z^{(xy,xy)}=-262.4$ meV is an inter-layer hopping parameter shown in Appendix~\ref{ap:model}.
Figures~\ref{fig:trgsw}(a)--(c) show the dispersion of the surface states by the solid lines and the continuous energy levels in the bulk by the shaded areas for the $\alpha$, $\beta$, and $\gamma$ bands, respectively.
The surface states at the momentum between the points where the SC gap closes are very close to zero energy \cite{Fukaya2}.
In addition, the surface states are not exactly at zero energy, as shown in Figs.~\ref{fig:trgsw}(d)--(f), when viewed at a higher energy resolution.
Parameters $\delta E_{\alpha,\beta,\gamma}$ in Figs.~\ref{fig:trgsw}(d), (e), and (f) show the width of the band splitting at each representative point on the $k_d a$ line for the $\alpha$, $\beta$, and $\gamma$ bands, respectively.
As shown in Fig. \ref{fig:split}, the $\Delta$ dependence of $\delta E_{\alpha,\beta,\gamma}$ is proportional to $\Delta^2$.
We demonstrate that the presence of nearly zero-energy surface states is due to the effective chiral $d$-wave-symmetry pair potential, whereas the tiny energy splitting proportional to $\Delta^2$ is inevitable because of the pseudo-Zeeman potential, as shown in the previous section.

\subsection{Surface local density of states} 
Second, we calculate momentum-resolved LDOS at $\varepsilon=0$, $-1/\pi \sum_{i}{\mathrm{Im}} \dot{g}^s_{ii}(\bm{k}_\parallel ,0)$. 
Figures \ref{fig:trgsk}(a), (e), and (i) show color plots of the momentum-resolved LDOS in the range shown in Fig. \ref{fig:FS} (c) for $\delta=\Delta\times10^{-2}$, $\Delta\times10^{-5}$, and $\Delta\times10^{-7}$, respectively.
Figures \ref{fig:trgsk}(b)--(d), (f)--(h), and (j)--(k) show enlarged maps surrounded by the squares in Figs. \ref{fig:trgsk}(a), (e), and (i), respectively.
The dotted and dashed lines show projected Fermi surface at $k_zc=0$ and $k_zc=2\pi$ onto the surface BZ, respectively.
For $\delta=\Delta\times10^{-2}$, the spectral weight
increases at the momenta between the dotted and dashed lines for the $\beta$ and $\gamma$ bands.
The spectral weight for the $\alpha$ band is weaker than those for the other bands.
One reason for this weak spectral weight might be the longer localization length of the surface state due to the smaller bulk energy gap in the momentum range enclosed by the dotted and dashed lines for this band.
Even outside the enclosed range, a weak spectral weight is observed near the dashed line.
This spectral weight arises from the inner-gap surface states outside of the enclosed momentum range, whose energy dispersion becomes closer to zero energy as the momenta approaches the dashed lines, as shown in Fig.~\ref{fig:trgsw}(a).
The $\gamma$ band also has the inner-gap surface states outside the enclosed momentum range, as shown in Fig.~\ref{fig:trgsw}(c).
However, these surface states are further from the zero energy than those of the $\alpha$ band.
When $\delta$ is set to a value close to the splitting energy scale shown in Figs.~\ref{fig:trgsw}(d)--(f), the obtained spectral weight inside the momentum range enclosed by the dotted and dashed lines clearly increases in each band in Figs.~\ref{fig:trgsk}(e)--(h).
For $\delta$ values much smaller than the splitting energy scale, the spectral weight at zero energy becomes weaker, as shown in Figs.~\ref{fig:trgsk}(i)--(l), because no surface states exist within the energy range less than $\delta$, as shown in Figs.~\ref{fig:trgsw}(d)--(f).

Figures~\ref{fig:DOS}(a) and (b) show the DOS of the bulk and the LDOS at the top surface, respectively, where we take $\Delta =|t_z^{(xy,xy)}\times10^{-4}|$ and $\delta=\Delta\times10^{-2}$. 
The DOS and the LDOS are normalized by those values at the Fermi energy in the normal state denoted by  $N_n$ and $N^s_n$, respectively.
As a result of the momentum-resolved LDOS for $\varepsilon=0$ shown in Fig. \ref{fig:trgsk}(a), a pronounced peak structure appears in the LDOS of the top surface at $\varepsilon=0$, in contrast to the V-shaped structure in the DOS of the bulk.
As we have explained, the surface state is not exactly at zero-energy. 
Therefore, if we take a sufficiently small $\delta$ to distinguish the surface state energy level shown in Fig.~\ref{fig:trgsw}(b), (d), and (f), the peak structure splits.
However, such a small resolution is challenging in actual experiments.

\subsection{$E_u$-symmetry perturbations}
\label{sec:discussion}
Finally,  we study the stability of the zero-energy peak of the LDOS against the perturbations due to inversion symmetry breaking, which is inevitable near the surface; such perturbations include orbital Rashba coupling that induces the Rashba-type spin--orbit coupling through the $L\cdot S$ coupling  \cite{Y-Yanase-SO,G-Khalsa} or an $E_u$ symmetry SC pair potential.
For the former case, we introduce the orbital Rashba coupling described by 
\begin{align}
  \hat{H}_{\mathrm{so}}=2t_{\mathrm{so}}
  (\bar{L}_y 
  \tilde{\sigma}_0 \sin(k_xa)+\bar{L}_x \tilde{\sigma}_0 \sin(k_ya)) 
\end{align}
at the top surface.
This coupling originates from the hopping integral due to the shift of the oxygen sites.
Thus, its value appears to be smaller than the nearest-neighbor hopping without the oxygen shift $t_x^{(xy,xy)}$, and we set $t_{\mathrm{so}}$ to approximately one-tenth  the value of $t_x^{(xy,xy)}$.
\begin{figure}[!htb]
  \includegraphics[width=1\linewidth]{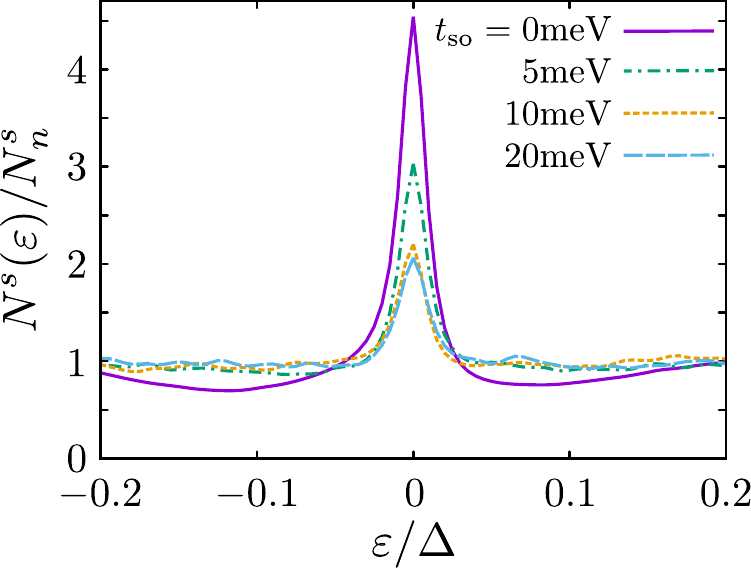}
  \caption{(Color online)
   The $t_{\mathrm{so}}$ dependence of the LDOS under the orbital Rashba coupling at the top surface.}
  \label{fig:additional_Rashba}
\end{figure}
The obtained LDOS is shown in Fig.~\ref{fig:additional_Rashba}.
As evident in this figure, the zero-energy peak remains under this range of $t_{\mathrm{so}}$. 
We confirm that the energy levels of the nearly zero-energy surface states do not change against this term.
However, its peak height is suppressed, possibly because the localized position of the surface state moves farther into the interior of the crystal. 
For the latter case, we introduce the chiral $E_u$ SC pair potential at the top surface instead of the $E_g$ inter-orbital SC pair potential.
The dependence of the LDOS on the pair potential and the gap amplitude are shown in Appendix \ref{ap:E_u}. 
Here, we set the amplitude of the gap as the order of $T_c$. 
In this case, the change of the resultant LDOS in Fig. \ref{fig:additional_Eu} is too small for them to be distinguished from the original ones given in Fig. \ref{fig:DOS}(b).   

\section{Discussion}
In this paper, we have only considered the clean surfaces preserving $\boldsymbol{k}_{\parallel}$ as a good quantum number.
However, in actual experiments, surface roughness will inevitably break translational symmetry.
When the surface roughness is substantial, for a $k_z(k_x+\mathrm{i}k_y)$-wave superconductor, the zero-energy surface states at the top surface are highly vulnerable~\cite{surfacce-roughness}.
In addition, in some experiments, surface reconstruction breaking four-fold rotation symmetry at the top surface of SRO has been observed ~\cite{A-Damascelli,R-Fittipaldi,C-A-Marques}.
Thus, examining the effect of surface roughness/reconstruction on the surface states of the present model would be an important future task.

In scanning tunneling microscopy (STM) observations of the top surface of SRO~\cite{R-Sharma,Firmo}, the differential conductance spectrum shows a V-shaped structure rather than a zero-bias peak structure, which appears to contradict the presence of nearly zero-energy surface states.
However, we emphasize that careful verification is needed to confirm whether these experimental data are indeed inconsistent with the realization of inter-orbital superconductivity in SRO.
Actually, a junction consisting of a \textit{one-dimensional} normal-metal lead wire and a two-dimensional $d_{xy}$-wave superconductor has been shown to exhibit a V-shaped conductance spectrum even in the presence of flat-band zero-energy surface states at the junction interface~\cite{takagaki_99,takagaki_01,takagaki_01(2),giraldo_05,Y-Tanaka,Y-Tanaka2,Y-Tanaka3}.
In general, this V-shaped spectrum is due to the sign change in the pair potential with respect to the momentum parallel to the surface; that is,
\begin{align}
\Delta_{d_{xy}}(k_x,k_y) = -\Delta_{d_{xy}}(k_x,-k_y),\nonumber
\end{align}
where $\Delta_{d_{xy}} \propto k_x k_y$ with $k_{x(y)}$ representing momentum perpendicular (parallel) to the interface.
More specifically, the Andreev reflection at the junction interface is substantially suppressed because of the destructive interference between the transmission channels feeling $\Delta_{d_{xy}}(k_x,k_y)$ and that feeling $\Delta_{d_{xy}}(k_x,-k_y)=-\Delta_{d_{xy}}(k_x,k_y)$.
In the present model, the effective pair potential satisfies
\begin{align}
\psi_{\nu}(\boldsymbol{k}_{\parallel},k_z) = -\psi_{\nu}(-\boldsymbol{k}_{\parallel},k_z),
\end{align}
where $\boldsymbol{k}_{\parallel}$ corresponds to the momentum parallel to the top surface.
Thus, as in the case of one-dimensional normal-metal/two-dimensional $d_{xy}$-wave superconductor junctions, we can expect that the Andreev reflection between the STM tip and the top surface of the SRO can be suppressed by a destructive interference effect.
Nevertheless, whether this effect and/or the surface perturbations (the surface reconstruction and the surface roughness) enables us to explain the absence of the zero-bias conductance peak in the STM data remains unclear.
To resolve this stalemate, microscopic studies on the transport properties of the  present model are strongly desired.

\section{Summary and Conclusion}\label{sec:summary}
We have studied the surface state of the inter-orbital-odd spin-triplet $s$-wave pairing at the (001) surface in SRO.
We have demonstrated the presence of nearly zero-energy surface states due to the effective chiral $d$-wave symmetry pair potential, whereas the tiny energy splitting proportional to $\Delta^2$ is inevitable because of the pseudo-Zeeman potential.
As a result, the surface LDOS  
exhibits a pronounced zero-energy peak when the order of the resolution is lower than the splitting energy shown in Fig.~\ref{fig:trgsw}(d)--(f).
We have also shown that the zero-energy peak is robust against perturbations due to inversion symmetry breaking at the surface, such as the orbital Rashba coupling and the $E_u$ symmetry pair potential.

\section{Acknowledgements}
This work was supported by Scientific Research (A) (KAKENHI Grant No. JP20H00131), and Scientific Research (B) (KAKENHI Grants No. JP20H01857).
S.I. is supported by a Grant-in-Aid for JSPS Fellows (JSPS KAKENHI Grant No. JP21J00041).
This work was supported by JSPS Core-to-Core Program (No. JPJSCCA20170002).
S.A. would like to take this opportunity to thank the “Nagoya University Interdisciplinary Frontier Fellowship” supported by Nagoya University and JST, the establishment of university fellowships towards the creation of science technology innovation, Grant Number JPMJFS2120.

\appendix
\setcounter{figure}{0}
\renewcommand{\thefigure}{A\arabic{figure}}
\renewcommand{\theequation}{A\arabic{equation}}
\section{
MODEL HAMILTONIAN OF THREE-DIMENSIONAL
 $\mathrm{Sr_2RuO_4}$ IN THE NORMAL STATE}
 \label{ap:model}
We describe the three-dimensional Hamiltonian of 
$\mathrm{Sr_2RuO_4}$ (SRO) in the normal state \cite{H-G-Suh}:
\begin{align} 
\hat{H}_{N}(\bm{k})
=
\sum_{lj}
h_{lj}(\bm{k})\bar{\lambda}_l
\otimes \tilde{\sigma}_j.
\end{align}
The Gell-Mann matrices $\bar{\lambda}_{l=0-8}$ are defined by  
\begin{align*} 
   &\bar{\lambda}_0=
   \left(
   \begin{array}{ccc}
        1               & 
        0               &
        0                \\
        0                &                          
        1                &
        0                \\
        0                & 
        0                &
        1                 \\                            
      \end{array}
    \right),
    \bar{\lambda}_1=
    \left(
    \begin{array}{ccc}
          0               & 
          1               &
          0                \\
          1                &                          
          0                &
          0                \\
          0                & 
          0                &
          0                 \\                            
        \end{array}
      \right),\\
      &\bar{\lambda}_2=
      \left(
      \begin{array}{ccc}
      0               & 
      0               &
      1                \\
      0                &                          
      0                &
      0                \\
      1                & 
      0                &
      0                 \\                            
      \end{array}
      \right),
      \bar{\lambda}_3=
      \left(
     \begin{array}{ccc}
      0               & 
      0               &
      0                \\
      0                &                          
      0                &
      1                \\
      0                & 
      1                &
      0                 \\                            
    \end{array}
  \right),\\
  &\bar{\lambda}_4=
\left(
\begin{array}{ccc}
      0               & 
      -\mathrm{i}               &
      0                \\
       \mathrm{i}                &                          
      0                &
      0                \\
      0                & 
      0                &
      0                 \\                            
    \end{array}
  \right),
  \bar{\lambda}_5=
\left(
\begin{array}{ccc}
      0               & 
      0               &
      -\mathrm{i}                \\
       0                &                          
      0                &
      0                \\
      \mathrm{i}                & 
      0                &
      0                 \\                            
    \end{array}
  \right),\\
  &\bar{\lambda}_6=
  \left(
  \begin{array}{ccc}
      0               & 
      0               &
      0                \\
       0                &                          
      0                &
      -\mathrm{i}                \\
      0               & 
      \mathrm{i}                &
      0                 \\                            
    \end{array}
  \right),
  \bar{\lambda}_7=
\left(
\begin{array}{ccc}
      1               & 
      0               &
      0                \\
       0                &                          
      -1                &
      0                \\
      0               & 
      0                &
      0                 \\                            
    \end{array}
  \right),\\
  &\bar{\lambda}_8=
\frac{1}{\sqrt{3}}\left(
\begin{array}{ccc}
      1               & 
      0               &
      0                \\
       0                &                          
      1                &
      0                \\
      0               & 
      0                &
      -2                 \\                            
    \end{array}
  \right).
  \end{align*}
 
$h_{lj}(\bm{k})$ are
    \begin{align}
        h_{00}(\bm{k})=\frac{1}{3}[\xi_{yz}(\bm{k})
                                  +\xi_{zx}(\bm{k})
                                  +\xi_{xy}(\bm{k})],
     \end{align}
        
        \begin{align}
          h_{70}(\bm{k})=\frac{1}{2}[\xi_{yz}(\bm{k})
                                    -\xi_{zx}(\bm{k})],
          \end{align}
        
        \begin{align}
            h_{80}(\bm{k})=\frac{1}{2\sqrt{3}}[\xi_{yz}(\bm{k})
                                      +\xi_{zx}(\bm{k})
                                      -2\xi_{xy}(\bm{k})],
        \end{align}

        \begin{align} 
          &h_{10}(\bm{k}) =g(\bm{k})\\
          =&-4t_{12xy}
          \sin{k_x}
          \sin{k_y} \nonumber\\
          &-4t_{12xxy}
          (\sin{2k_x}
          \sin{k_y}+
          \sin{k_x}
          \sin{2k_y})\nonumber\\
          &+8t_z^{z}\sin{\frac{k_xa}{2}}\sin{\frac{k_ya}{2}}\cos{\frac{k_zc}{2}},
          \end{align}
          
\begin{align} 
h_{20}(\bm{k})=8t_{z}^{(zx,xy)} 
\sin{\frac{k_xa}{2}}\sin{\frac{k_ya}{2}}\cos{\frac{k_zc}{2}},
\end{align}          
      
\begin{align} 
h_{30}(\bm{k})=8t_{z}^{(zx,xy)} 
\sin{\frac{k_xa}{2}}\cos{\frac{k_ya}{2}}\sin{\frac{k_zc}{2}},
\end{align}    
     
\begin{align} 
h_{43}(\bm{k})=-\lambda^{\mathrm{SOC}},
\end{align}

\begin{align} 
h_{52}(\bm{k})=
-h_{61}(\bm{k})=
\lambda^{\mathrm{SOC}},
\end{align}

\begin{align} 
h_{51}(\bm{k})=
-h_{62}(\bm{k})=
4\lambda_{5162}^{\mathrm{SOC}},
\sin{k_xa}\sin{k_ya}
\end{align}

\begin{align} 
h_{52}(\bm{k})=
h_{61}(\bm{k})=
2\lambda^{\mathrm{SOC}}_{5261}(\cos{k_xa}-\cos{k_ya}),
\end{align}

\begin{align} 
h_{41}(\bm{k})=8\lambda_{12z}^{\mathrm{SOC}} 
\sin{\frac{k_xa}{2}}\sin{\frac{k_ya}{2}}\cos{\frac{k_zc}{2}},
\end{align}        
       
\begin{align} 
h_{42}(\bm{k})=8\lambda_{12z}^{\mathrm{SOC}} 
\sin{\frac{k_xa}{2}}\cos{\frac{k_ya}{2}}\sin{\frac{k_zc}{2}},
\end{align}

\begin{align} 
h_{63}(\bm{k})=-8\lambda_{56z}^{\mathrm{SOC}} 
\sin{\frac{k_xa}{2}}\sin{\frac{k_ya}{2}}\cos{\frac{k_zc}{2}},
\end{align}         
          
\begin{align} 
h_{53}(\bm{k})=8\lambda_{56z}^{\mathrm{SOC}} 
\sin{\frac{k_xa}{2}}\cos{\frac{k_ya}{2}}\sin{\frac{k_zc}{2}},
\end{align}
 with         
\begin{align} 
    \xi_{yz}=
    &2t_y^{(z,z)}\cos{k_x a}
    +2t_x^{(z,z)}\cos{k_ya}-\mu_z-\delta\mu \notag \\&
    +4t_{xy}^{(z,z)}\cos{k_xa}\cos{k_ya} \notag \\&
    +2t_{yy}^{(z,z)}\cos{2k_xa}
    +2t_{xx}^{(z,z)}\cos{2k_ya} \notag \\&
    +4t_{xyy}^{(z,z)}\cos{2k_xa}\cos{k_ya}
    +4t_{xxy}^{(z,z)}\cos{2k_ya}\cos{k_xa} \notag \\&
    +8t_z^{(z,z)}\cos{\frac{k_xa}{2}}\cos{\frac{k_ya}{2}}\cos{\frac{k_zc}{2}} \notag \\&
    +2t_{zz}^{(z,z)}
    (\cos{k_zc}-1),
  \end{align}

  \begin{align} 
      \xi_{zx}=
      &2t_x^{(z,z)}\cos{k_x a}
      +2t_y^{(z,z)}\cos{k_ya}-\mu_z-\delta\mu \notag \\&
      +4t_{xy}^{(z,z)}\cos{k_xa}\cos{k_ya} \notag \\&
      +2t_{xx}^{(z,z)}\cos{2k_xa}
      +2t_{yy}^{(z,z)}\cos{2k_ya} \notag \\&
      +4t_{xxy}^{(z,z)}\cos{2k_xa}\cos{k_ya}
      +4t_{xyy}^{(z,z)}\cos{2k_ya}\cos{k_xa} \notag \\&
      +8t_z^{(z,z)}\cos{\frac{k_xa}{2}}\cos{\frac{k_ya}{2}}\cos{\frac{k_zc}{2}}\notag \\&
      +2t_{zz}^{(z,z)}(\cos{k_zc}-1),
   \end{align} 

    \begin{align}  
        \xi_{xy}=\
        &2t_x^{(xy,xy)}(\cos{k_xa}+\cos{k_ya})-\mu_{xy}-\delta\mu \notag \\&
        +4t_{xy}^{(xy,xy)}\cos{k_xa}\cos{k_ya} \notag \\&
        +2t_{xx}^{(xy,xy)}(\cos{2k_ya}+\cos{2k_xa}) \notag \\&
        +4t_{xxy}^{(xy,xy)}
        (\cos{2k_xa}\cos{k_ya}+\cos{k_xa}\cos{2k_ya}) \notag \\&
        +8t_z^{(xy,xy)}\cos{\frac{k_xa}{2}}\cos{\frac{k_ya}{2}}\cos{\frac{k_zc}{2}} \notag \\&
        +2t_{zz}^{(xy,xy)}(\cos{2k_zc}-1).
    \end{align}

We set the parameters as shown in Table \Ref{table:parameters} \cite{H-G-Suh}.

    \begin{table}[hbtp]
        \caption{Parameters for the normal state of the Hamiltonian. The unit is meV.}
        \label{table:parameters}
        \centering
        \begin{tabular}{clll}
             \hline \hline  
             & $t_{x}^{(z,z)}=-362.4$ \quad$t_{y}^{(z,z)}=-134$
             \quad$t_{x}^{(xy,xy)}=-262.4$\\             
             &$t_{xy}^{(z,z)}=-44.01$\quad              
            $t_{xx}^{(z,z)}=-1.021$\quad $t_{yy}^{(z,z)}=-5.727$\\
             &$t_{xx}^{(xy,xy)}=34.23$ \quad$t_{xy}^{z}=16.25$\quad
             $t_{xxy}^{(z,z)}=-13.93$\\ 
             &$t_{xyy}^{(z,z)}=-7.52$ 
             \quad$t_{xxy}^{(xy,xy)}=8.069$ \quad$t_{xxy}^{z}=3.94$\\
             &$\lambda_{\mathrm{SO}}=57.39$\quad $\mu_z=438.5$ \quad 
              $\mu_{xy}=218.6$\\
            &$t_{z}^{(z,z)}=-0.0228$\quad
            $t_{z}^{(xy,xy)}=1.811$ \quad$t_{z}^{(z,z)}=9.975$ \\
             &$t_{z}^{(zx,xy)}=8.304$\quad$t_{zz}^{(z,z)}=2.522$\quad
            $t_{zz}^{(xy,xy)}=-3.159$\\
             &$\lambda^{\mathrm{SOC}}_{56z}=-1.247$\quad$\lambda^{\mathrm{SOC}}_{12z}=-3.576$
             \quad$\lambda^{\mathrm{SOC}}_{5162}=-1.008$ \\
             &\quad $\lambda^{\mathrm{SOC}}_{5261}=0.3779$\\
             \hline
        \end{tabular}
    \end{table}


\setcounter{figure}{0}
\renewcommand{\thefigure}{B\arabic{figure}}
\renewcommand{\theequation}{B\arabic{equation}}    
\section{EFFECTIVE LOW ENERGY MODEL}
\label{ap:low-energy}

\begin{figure*}[htbp]
\includegraphics[width=1\linewidth]{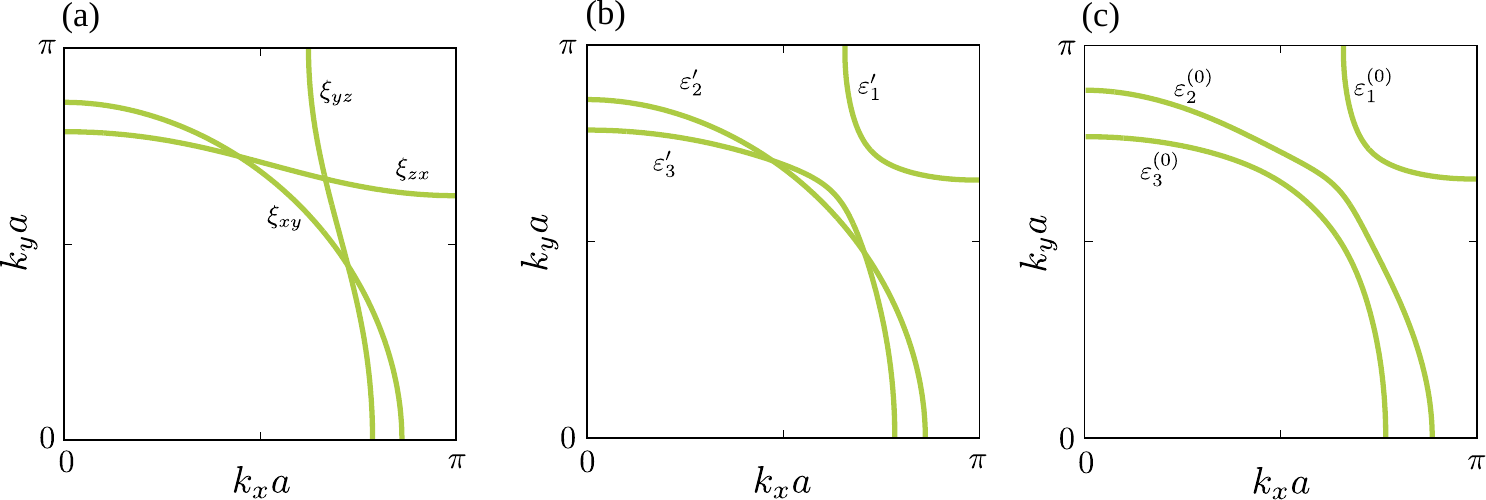}
\caption{(Color online)
(a) $\xi_{yz}$, $\xi_{zx}$, $\xi_{xy}$ at $k_z=0$ at the Fermi energy. 
(b) $\varepsilon_1'$, $\varepsilon_2'$, and $\varepsilon_3'$ at $k_z=0$ at the Fermi energy.
(c) $\varepsilon^{(0)}_1$, $\varepsilon^{(0)}_2$, and $\varepsilon^{(0)}_3$ at $k_z=0$ at the  Fermi energy.
}
\label{fig:FS_u}
\end{figure*}
In this Appendix, we derive a low-energy Hamiltonian of SRO given by Eq.~(\ref{eq:BdG}) in the main text. For this purpose, we first deform the BdG Hamiltonian in an approximate band-basis (see Eq.~(\ref{eq:band_ham}) in the main text). Then, we construct a low-energy effective Hamiltonian for each band by following Ref.~\cite{P-M-R-Brydon,J-W-F-Venderbos}.

In principle, the Hamiltonian in the band basis is obtained by employing a unitary operator that diagonalizes the normal Hamiltonian $\hat{H}_N$. 
In this paper, owing to the difficulty of the exact diagonalization, we alternatively utilize a perturbation theory to diagonalize $\hat{H}_N$ approximately.
As a preliminary step, we  remove all degeneracy in the diagonal part of $\hat{H}_N$ at the Fermi level as shown in Fig. ~\ref{fig:FS_u}(a). 
We now define a unitary operator,
\begin{align}
    \hat{u}_p=\hat{u}_1\hat{u}_2.
\end{align}
The first component of the unitary operator, i.e., $\hat{u}_1$, is defined by
\begin{align}
    \hat{u}_1=  \left(
        \begin{array}{cc}
           \bar{u}_1 & 0 \\
              0 & \bar{u}_1^*
            \end{array}
          \right),
  \end{align}   
\begin{align}
    \bar{u}_1=\left(
        \begin{array}{ccc}
          a_1 & -b^*_1 & 0 
          \\
          b_1 & a_1 & 0
          \\
          0 & 0 & 1
        \end{array}
      \right) ,       
\end{align}
with
\begin{align}
    &a_1=\frac{P_1+R_1}{\sqrt{2P_1(P_1+R_1)}},\\
    &b_1=\frac{h_{10}+\mathrm{i}h_{43}}{\sqrt{2P_1(P_1+R_1)}},\\    
    &R_1=(\xi_{yz}+\xi_{zx})/2,\\
    &P_1=\sqrt{{R_1}^2+{h_{10}}^2+{h_{43}}^2}.
\end{align}
By applying $\hat{u}_1$ to $\hat{H}_N$ , we obtain
\begin{align}
\hat{H}^\prime_N=\hat{u}_1^{\dag}\hat{H}_N\hat{u}_1.    
\end{align}
The diagonal components of $\hat{H}_N^{\prime} $ are given by
\begin{align}
    &\varepsilon_1' =Q_1+P_1,\\
    & \varepsilon_2' =Q_1-P_1,\\
    & \varepsilon_3'=\xi_{yz},
\end{align}  
with
\begin{align}
    Q_1=(\xi_{yz}-\xi_{zx})/2.
\end{align}
By using the unitary transformation of $\hat{u}_1$, the degeneracy between $\xi_{yz}$ and $\xi_{zx}$ at the Fermi level is lifted as shown in Fig.~\ref{fig:FS_u}(a). 
The second component of $\hat{u}_p$, i.e., $\hat{u}_2$, is defined by  
\begin{align}
    \hat{u}_2=  \left(
        \begin{array}{cc}
           \bar{a}_2 & -\bar{b}_2^* \\
              \bar{b}_2 & \bar{a}_2
            \end{array}
          \right),
  \end{align}   
\begin{align}
    \bar{a}_2=\left(
        \begin{array}{ccc}
          a_2 & 0 & 0 
          \\
          0 & 1 & 0
          \\
          0 & 0 & a_2
        \end{array}
      \right) ,       
  \end{align}
  
  \begin{align}
    \bar{b}_2=\left(
        \begin{array}{ccc}
          0 & 0 & b_2 
          \\
          0 & 1 & 0
          \\
          b_2 & 0 & 0
        \end{array}
      \right) ,       
  \end{align}
 with

\begin{align}
  &a_2=\frac{P_2+R_2}{\sqrt{2P_2(P_2+R_2)}},\\
  &b_2=\frac{-g_{52}+\mathrm{i}g_{51}}{\sqrt{2P_1(P_1+R_1)}},\\    
  &R_2=(\varepsilon_1'-\varepsilon_3')/2,\\
  &P_2=\sqrt{{R_2}^2+{g_{51}}^2+{g_{52}}^2},\\
  &g_{51} =\frac{(P_1+R_1)h_{51}+h_{10}h_{61}-h_{43}h_{62}}{\sqrt{2P_1(P_1+R_1)}},\\
  &g_{52} =\frac{(P_1+R_1)h_{52}+h_{10}h_{62}+h_{43}h_{61}}{\sqrt{2P_1(P_1+R_1)}}.
\end{align}
By applying $\hat{u}_2$ to $\hat{H}^\prime_N$, we obtain
\begin{align}
\hat{h}_p=\hat{u}_2^{\dag}\hat{H}_N^\prime\hat{u}_2=\hat{\varepsilon}^{(0)}+\hat{\eta},
\end{align}
with
\begin{align}
    \hat{\varepsilon}^{(0)}
    =   \left(
    \begin{array}{cc}
       \bar{\varepsilon}^{(0)} & 0 \\
          0   &  \bar{\varepsilon}^{(0)}
        \end{array}
      \right),
\end{align}
\begin{align}
    \hat{\eta}=
     \left(
     \begin{array}{cc}    
        \bar{\eta}_{0} +\bar{\eta}_{3}& \bar{\eta}_{1}-i\bar{\eta}_{2} 
         \\
         \bar{\eta}_{1}+ i\bar{\eta}_{2}& \bar{\eta}_{0}-\bar{\eta}_{3}
       \end{array}
     \right),
 \end{align}
 with
 \begin{align}
    \bar{\varepsilon}^{(0)}=
    \left(
    \begin{array}{ccc}
        {\varepsilon}^{(0)}_{1} & 0 & 0 
        \\
        0 & {\varepsilon}^{(0)}_{2} & 0
        \\
        0 & 0 & {\varepsilon}^{(0)}_{3}
      \end{array}
    \right), 
\end{align}
\begin{align}
\label{eq:eta_0}
  \bar{\eta}_0=
  \left(
   \begin{array}{ccc}
      0 & \eta_{10} & \eta_{20} 
      \\
      \eta_{10} & 0 & \eta_{30}
      \\
      \eta_{20} & \eta_{30} & 0
   \end{array}
 \right),
\end{align}
\begin{align}
\label{eq:eta_3}
    \bar{\eta}_3=
    \left(
     \begin{array}{ccc}
        0 & -\mathrm{i}\eta_{43} & -\mathrm{i}\eta_{53} 
        \\
        \mathrm{i}\eta_{43} & 0 & -\mathrm{i}\eta_{63}
        \\
        \mathrm{i}\eta_{53} & \mathrm{i}\eta_{63} & 0
     \end{array}
   \right),
\end{align}
\begin{align}
\label{eq:eta_12}
    \bar{\eta}_{j=1,2}=
    \left(
     \begin{array}{ccc}
        0 & -\mathrm{i}\eta_{4j} & 0 
        \\
        \mathrm{i}\eta_{4j} & 0 & -\mathrm{i}\eta_{6j}
        \\
        0 & \mathrm{i}\eta_{6j} & 0
     \end{array}
   \right).
  \end{align}    
The diagonal components of $\hat{h}_p$
is given by
\begin{align}
{\varepsilon}_{1}^{(0)}=Q_2+P_2,\\
    {\varepsilon}_{2}^{(0)}=Q_1-P_1,\\
    {\varepsilon}_{3}^{(0)}=Q_2-P_2,
\end{align}
with
\begin{align}
    &Q_2=(\varepsilon_1'+\varepsilon_3')/2.
\end{align}
The remained degeneracy between
$\varepsilon_{2}^\prime$ and $\varepsilon_{3}^\prime$ is lifted by $\hat{u}_2$ as shown in Fig.~\ref{fig:FS_u}(b).
The off-diagonal components of $\hat{h}_p$
is given by 
\begin{align}
\eta_{10}=\frac{g_{51}g_{61}+g_{52}g_{62}}{\sqrt{2P_2(P_2+R_2)}},
\end{align}
\begin{align}
  \eta_{20}=g_{20},
\end{align}
\begin{align}
  \eta_{30}=\frac{(P_2+R_2)g_{30}+g_{51}g_{41}-g_{52}g_{42}}{\sqrt{2P_2(P_2+R_2)}},
\end{align}
\begin{align}
    &\eta_{43}=\frac{-g_{51}g_{62}+g_{52}g_{61}}{\sqrt{2P_2(P_2+R_2)}},\\
&\eta_{53}=g_{53},\\
&\eta_{63}=\frac{(P_2+R_2)g_{30}+g_{51}g_{41}-g_{52}g_{42}}{\sqrt{2P_2(P_2+R_2)}},
\end{align}
\begin{align}
       & \eta_{41}=\frac{(P_2+R_2)g_{41}-g_{51}g_{30}-g_{52}g_{63}}{\sqrt{2P_2(P_2+R_2)}},\\
        &\eta_{42}=\frac{(P_2+R_2)g_{42}-g_{51}g_{63}-g_{52}g_{30}}{\sqrt{2P_2(P_2+R_2)}},\\
        &\eta_{61}=\frac{(P_2+R_2)g_{61}}{\sqrt{2P_2(P_2+R_2)}},\\
        &\eta_{62}=\frac{(P_2+R_2)g_{62}}{\sqrt{2P_2(P_2+R_2)}},  
\end{align}
with
\begin{align}
  &g_{20} =\frac{(P_1+R_1)h_{20}-h_{10}h_{30}-h_{43}h_{63}}{\sqrt{2P_1(P_1+R_1)}},\\
  &g_{30} =\frac{(P_1+R_1)h_{30}-h_{10}h_{20}-h_{43}h_{53}}{\sqrt{2P_1(P_1+R_1)}},\\
  &g_{4j}=h_{4j},\\
  &g_{53} =\frac{(P_1+R_1)h_{53}+h_{10}h_{63}+h_{43}h_{30}}{\sqrt{2P_1(P_1+R_1)}},\\
  &g_{63} =\frac{(P_1+R_1)h_{63}-h_{10}h_{53}+h_{43}h_{20}}{\sqrt{2P_1(P_1+R_1)}},\\
  &g_{61} =\frac{(P_1+R_1)h_{61}-h_{10}h_{51}-h_{43}h_{52}}{\sqrt{2P_1(P_1+R_1)}},\\
  &g_{62} =\frac{(P_1+R_1)h_{62}-h_{10}h_{52}+h_{43}h_{51}}{\sqrt{2P_1(P_1+R_1)}}.    
\end{align}
In short, 
all degeneracy at the Fermi level is lifted by the unitary transformation of $\hat{u}_p$, i.e.,
\begin{align}
\hat{u}_p^{\dag}\hat{H}_N\hat{u}_p=\hat{h}_p=\hat{\varepsilon}^{(0)}+\hat{\eta}.
\end{align}
For the BdG Hamiltonian, the corresponding unitary transformation is represented by
\begin{align}
\check{H}_p&=\check{u}_p^{\dag}\check{H}\check{u}_p\nonumber\\&=\left(
        \begin{array}{cc}
          \hat{h}_p(\bm{k}) & \hat{\Delta}_p(\bm{k}) 
          \\
          \hat{\Delta}_p^{\dag}(\bm{k}) & -\hat{h}_p^{\mathrm{T}}(-\bm{k}) 
        \end{array}
      \right)       
\end{align}
with
\begin{align}
\check{u}_p=\check{u}_1 \check{u}_2    
\end{align}
\begin{align}
    \check{u}_1=  \left(
        \begin{array}{cc}
           \hat{u}_1 & 0 \\
              0 & \hat{u}_1^*
            \end{array}
          \right),
\end{align}   
\begin{align}
    \check{u}_2=  \left(
        \begin{array}{cc}
           \hat{u}_2 & 0 \\
              0 & \hat{u}_2^*
            \end{array}
          \right),
\end{align}   
where the pair potential in this basis is given as
\begin{align}
    \hat{\Delta}_p=
    \left(
     \begin{array}{cc}
        -\bar{f}_1+\mathrm{i}\bar{f}_2 & \bar{f}_3+\bar{f}_0 
        \\
        \bar{f}_3-\bar{f}_0    & \bar{f}_1+\mathrm{i}\bar{f}_2 
     \end{array}
   \right),
\end{align}
with
\begin{align}
    \bar{f}_{3}=
    \left(
     \begin{array}{ccc}
        0 & 0 & {f}_{3,13}
        \\
        0 & 0 & \mathrm{i}{f}_{3,23}
        \\
        -{f}_{3,13} & -\mathrm{i}{f}_{3,23} & 0
     \end{array}
   \right),
\end{align}
\begin{align}
    \bar{f}_{0}=
    \left(
     \begin{array}{ccc}
        0 & 0 & {f}_{0,13}
        \\
        0 & 0 & -\mathrm{i}{f}_{0,23}
        \\
        \bar{f}_{0,13} & -\mathrm{i}{f}_{0,23} & 0
     \end{array}
   \right),
  \end{align}
  \begin{align}
    \bar{f}_{j=1,2}=
    \left(
     \begin{array}{ccc}
        0 & {f}_{j,12} & 0
        \\
        -{f}_{j,12} & 0 & 0
        \\
        0 & 0 & 0
     \end{array}
   \right),
\end{align}
with
\begin{align} 
  &f_{j,12}=\Delta(\phi_{j,12}+\mathrm{i}\chi_{j,12}),\\
 &f_{3,13}=\Delta(\phi_{3,13}+\mathrm{i}\chi_{3,13}),\\
 &f_{3,23}=\Delta(\phi_{3,23}+\mathrm{i}\chi_{3,23}),\\
  &f_{0,13}=\Delta \phi_{0,13},\\
  &f_{0,23}=\Delta \phi_{0,23},    
   \end{align}
with 
 \begin{align}
       &\phi_{1,12}=\frac{h_{10}g_{52}-h_{43}g_{51}}{\sqrt{2P_1(P_1+R_1)}\sqrt{2P_2(P_2+R_2)}},\\
      &\chi_{1,12}=\frac{(P_1+R_1)g_{52}}{\sqrt{2P_1(P_1+R_1)}\sqrt{2P_2(P_2+R_2)}},\\
      &\phi_{2,12}=\frac{-h_{10}g_{51}-h_{42}g_{52}}{\sqrt{2P_1(P_1+R_1)}\sqrt{2P_2(P_2+R_2)}},\\
      &\chi_{2,12}=\frac{(P_1+R_1)g_{51}}{\sqrt{2P_1(P_1+R_1)}\sqrt{2P_2(P_2+R_2)}},\\
      &\phi_{3,13}=\frac{P_1+R_1}{\sqrt{2P_1(P_1+R_1)}\sqrt{2P_2(P_2+R_2)}},\\
      &\chi_{3,13}=\frac{(P_1+R_1)g_{52}}{\sqrt{2P_1(P_1+R_1)}\sqrt{2P_2(P_2+R_2)}},\\     
      &\phi_{3,23}=\frac{(P_1+R_1)(P_2+R_2)}{\sqrt{2P_1(P_1+R_1)(P_2+R_2)}\sqrt{2P_2(P_2+R_2)}},\\
      &\chi_{3,23}=\frac{(P_2+R_2)h_{10}}{\sqrt{2P_1(P_1+R_1)}\sqrt{2P_2(P_2+R_2)}},\\
      &\phi_{0,13}=\frac{h_{43}}{\sqrt{2P_1(P_1+R_1)}},\\
      &\phi_{0,23}=\frac{(P_2+R_2)h_{43}}{
      \sqrt{2P_1(P_1+R_1)}
      \sqrt{2P_2(P_2+R_2)}}.
 \end{align} 
For later convenience, we additionally apply a unitary transformation:
\begin{align}
    \check{\mathcal{H}}_q=\check{u}_q^{\dag} \check{\mathcal{H}}_p \check{u}_q=
    \left(
     \begin{array}{cc}
        \hat{h}_q(\bm{k}) & \hat{\Delta}_q(\bm{k}) 
        \\
        \hat{\Delta}_q^{\dag}(\bm{k}) & -\hat{h}_q^{\mathrm{T}}(-\bm{k}) 
     \end{array}
     \right),
\end{align} 
with
\begin{align}
\check{u}_q=\left(
    \begin{array}{cc} 
        \hat{u}_q& 0 
        \\
        0 & \hat{u}_q 
          \end{array}
    \right), 
\end{align}
\begin{align}
\hat{u}_q=    
\left(
    \begin{array}{cccccc}
        1& 0 & 0 & 0 & 0& 0
        \\
        0 & 0 & 0& 0 & 1& 0
        \\
        0 & 0 & 1& 0 & 0& 0
        \\
        0 & 1 & 0& 0 & 0& 0
        \\
        0 & 0 & 0& 1 & 0& 0
        \\
        0 & 0 & 0& 0 & 0& 1
      \end{array}
    \right). 
\end{align}
For the normal part,
\begin{align}
    \hat{h}_q= \hat{\varepsilon}_q^{(0)}+\hat{\eta}_q,
\end{align}
with
\begin{align}
    \hat{\varepsilon}^{(0)}_q=
    \left(
    \begin{array}{ccc}
        \varepsilon_{1}^{(0)} \tilde{\sigma}_0& 0 & 0 
        \\
        0 & \varepsilon_{2}^{(0)}\tilde{\sigma}_0 & 0
        \\
        0 & 0 & \varepsilon_{3}^{(0)}\tilde{\sigma}_0
      \end{array}
    \right), 
\end{align}
\begin{align}
    \hat{\eta}_q=
    \left(
     \begin{array}{ccc}
        0 & \hat{\eta}_{12} & \hat{\eta}_{13}
        \\
        \hat{\eta}_{21} & 0 & \hat{\eta}_{23}
        \\
        \hat{\eta}_{31} & \hat{\eta}_{32} & 0
     \end{array}
   \right),
\end{align}
with
\begin{align}
    \hat{\eta}_{ij}=
    \left(
    \begin{array}{cc}
        \eta_{0,ij}+\mathrm{i}\eta_{3,ij} & \eta_{1,ij}-\mathrm{i}\eta_{2,ij}  
        \\
        \eta_{1,ij}+\mathrm{i}\eta_{2,ij} & \eta_{0,ij}-\mathrm{i}\eta_{3,ij} 
     \end{array}
     \right),
\end{align}
where
\begin{align}
    \eta_{\alpha,ij}=(\bar{\eta}_\alpha)_{ij}.
\end{align}
For the pair potential part,
\begin{align}
    \hat\Delta_q=
    \left(
     \begin{array}{ccc}
        0 & \hat{\Delta}_{12} & \hat{\Delta}_{13}
        \\
        \hat{\Delta}_{21} & 0 & \hat{\Delta}_{23}
        \\
        \hat{\Delta}_{31} & \hat{\Delta}_{32} & 0
     \end{array}
   \right),
\end{align}
with
\begin{align}
    \hat{\Delta}_{ij}=
    \left(
    \begin{array}{cc}
        -\Delta_{1,ij}+\mathrm{i}\Delta_{2,ij} & \Delta_{3,ij}+\Delta_{0,ij}  
        \\
        \Delta_{3,ij}-\Delta_{0,ij} & \Delta_{1,ij}+\mathrm{i}\Delta_{2,ij} 
     \end{array}
     \right),
\end{align}
where
\begin{align}
    \Delta_{\alpha,ij}=(\bar{f}_{\alpha})_{ij}.
\end{align}

To proceed the approximate diagonalization of $\hat{H}_N$, we now consider a perturbation theory with assuming $\hat{\eta}$ as a perturbation. The unperturbed term $\hat{\varepsilon}^{(0)}$ satisfies 
\begin{align}
    \begin{split}
    \hat{u}_1^{(0)}=
    \left(
     \begin{array}{ccc}
        \tilde{\sigma}_0 & 0 & 0
        \\
        0 & 0 & 0
        \\
        0 & 0 & 0
     \end{array}
   \right),
   \hat{u}_2^{(0)}=
   \left(
    \begin{array}{ccc}
       0 & 0 & 0
       \\
       0 & \tilde{\sigma}_0 & 0
       \\
       0 & 0 & 0
    \end{array}
  \right),\\
  \hat{u}_3^{(0)}=
  \left(
    \begin{array}{ccc}
       0 & 0 & 0
       \\
       0 & 0 & 0
       \\
       0 & 0 & \tilde{\sigma}_0
    \end{array}
  \right),
\end{split}
\end{align}
\begin{align}
\hat{\varepsilon}^{(0)}\hat{u}_j^{(0)}=\varepsilon_j^{(0)}\hat{u}^{(0)}_j.
\end{align}
where $\varepsilon^{(0)}_j$ for $j=1,2,3$ have no degeneracy at the Fermi level. Thus, on the basis of the perturbation theory, we can diagonalize the perturbed Hamiltonian $\hat{h}_q$ within the second order of $\hat{\eta}$ as 
 \begin{align}
&\hat{\varepsilon}=\hat{u}_r^{\dag} \hat{h}_q
\hat{u}_r\nonumber\\
    &=\left(
      \begin{array}{ccc}(\varepsilon_1^{(0)}+\mu_{\lambda,1})\tilde{\sigma}_0 & 0 & 0
         \\
         0 & (\varepsilon_2^{(0)}+\mu_{\lambda,2})\tilde{\sigma}_0 & 0
         \\
         0 & 0 & (\varepsilon_3^{(0)}+\mu_{\lambda,3})\tilde{\sigma}_0
      \end{array}
    \right)\nonumber\\
    &+\mathcal{O}(\eta^3),
    \label{eq:aa}
    \end{align}
where
\begin{align}
    \mu_{\lambda,i}=\sum_{k\neq i}\frac{|\eta_{0,ki}|^2+|\bm{\eta}_{ki}|^2}{(\varepsilon_i-\varepsilon_k)}.
\end{align}
The last term of Eq.~(\ref{eq:aa}), i.e., $O({\eta}^n)$, represents the Landau symbol with respect to the $n$-th order of the matrix elements in $\hat{\eta}_q$.
The unitary transformation $\hat{u}_r$ given by
\begin{align}
\hat{u}_r&=\hat{u}_1+\hat{u}_2+\hat{u}_3\nonumber\\
        &=\hat{1}+\hat{W}+\hat{Z},         
\end{align}
with
\begin{align}
        \hat{W}=
        \left(
          \begin{array}{ccc}
             0 & \hat{W}_{12} & \hat{W}_{13}
             \\
             \hat{W}_{21} & 0 & \hat{W}_{23}
             \\
             \hat{W}_{31} & \hat{W}_{32} & 0
          \end{array}
        \right),
\end{align}
\begin{align}
        \hat{Z}=
        \left(
          \begin{array}{ccc}
            \hat{Z}_{11} & \hat{Z}_{12} & \hat{Z}_{13}
             \\
             \hat{Z}_{21} & \hat{Z}_{22} & \hat{Z}_{23}
             \\
             \hat{Z}_{31} & \hat{Z}_{32} & \hat{Z}_{33}
          \end{array}
        \right),
\end{align}
where
\begin{align}
    &\hat{W}_{ij}=\frac{\hat{\eta}_{ij}}{\varepsilon_j-\varepsilon_i},\\
       & \hat{Z}_{ii}=-\frac{1}{2}\sum_{j\neq i}\frac{\hat{\eta}_{ji}^{\dag}\hat{\eta}_{ji}}
        {(\varepsilon_j-\varepsilon_i)^2},\\
        &\hat{Z}_{ij}=\sum_{l \neq j,i}\frac{\hat{\eta}_{il}\hat{\eta}_{lj}}{(\varepsilon_j-\varepsilon_i)(\varepsilon_j-\varepsilon_l)}.
\end{align}
As a result, we can deform the BdG Hamiltonian in the approximate band basis as
\begin{align}
\check{\mathcal{H}}_r=&
\check{u}_r^{\dag}\check{\mathcal{H}}_q
\check{u}_r 
=
  \left(
    \begin{array}{cc}
       \hat{\varepsilon}(\bm{k}) & \hat{\mathcal{D}} (\bm{k})\\
       \hat{\mathcal{D}}^{\dag} (\bm{k})  &-\hat{\varepsilon}^{\mathrm{T}}(-\bm{k}) 
    \end{array}
  \right),
\end{align}
with
\begin{align}
    \check{u}_r=
    \left(
          \begin{array}{cc}
             \hat{u}_r & 0 
             \\
             0 & \hat{u}_r^* 
          \end{array}
        \right),
\end{align}
where
\begin{align} 
   &\hat{\mathcal{D}}
    =
    \left(
      \begin{array}{ccc}
        \hat{\mathcal{D}}_{11} & \hat{\mathcal{D}}_{12} & \hat{\mathcal{D}}_{13}
         \\
         \hat{\mathcal{D}}_{21} & \hat{\mathcal{D}}_{22} & \hat{\mathcal{D}}_{23}
         \\
         \hat{\mathcal{D}}_{31} & \hat{\mathcal{D}}_{32} & \hat{\mathcal{D}}_{33}
      \end{array}
    \right),
\end{align}
with
\begin{align}
\hat{\mathcal{D}}_{ij}&=\hat{\Delta}_{ij}\nonumber\\
   &+\sum_{k\neq i,j}
   (-\hat{W}_{ik}\hat{\Delta}_{kj}+\hat{\Delta}_{ik}\hat{W}_{kj}^*)+\mathcal{O}({\eta}^2).
\end{align}
 We can expand $\hat{D}_{ij}$ as
\begin{align}
    &\hat{\mathcal{D}}_{ij}=  
    \left(
        \begin{array}{cc}
          d_{1,ij}-\mathrm{i}d_{2,ij} &  -d_{3,ij}+\psi_{ij}
           \\
           -d_{3,ij}-\psi_{ij} & -d_{1,ij}-\mathrm{i}d_{2,ij}
        \end{array}
      \right),
\end{align}
where $d_{\alpha,ij}$ are the inter-band spin-triplet pair potential, and 
$\psi_{ij}$ are the inter- and intra-band  spin-singlet pair potential.  
It should be noted that intra-band pair potential is composed only of the spin-singlet pair potential, i.e.,  
\begin{align}
    &\hat{\mathcal{D}}_{ii}=  
    \left(
        \begin{array}{cc}
          0 &  \psi_{ii}
           \\
           -\psi_{ii} & 0
        \end{array}
      \right),
\end{align}
where 
\begin{align}
    \begin{split}
        \psi_{11}=&2\Delta\left( -\frac{\eta_{41}\chi_{1,12}+\eta_{42}\chi_{2,12}}{\varepsilon_1-\varepsilon_2} \right.\\ &\left.
        -\frac{\eta_{20}\phi_{0,13}-\eta_{53}\chi_{3,13}}{\varepsilon_3-\varepsilon_1} \right)\\
        &+\mathrm{i}2\Delta\left( \frac{\eta_{41}\phi_{1,12}+\eta_{42}\phi_{2,12}}{\varepsilon_1-\varepsilon_2} 
        -\frac{\eta_{53}\phi_{3,13}}{\varepsilon_3-\varepsilon_1} \right),\\   
        \psi_{22}=&2\Delta  \left( \frac{\eta_{41}\chi_{1,12}+\eta_{42}\chi_{1,12}}{\varepsilon_1-\varepsilon_2} 
        -\frac{\eta_{63}\phi_{3,23}}{\varepsilon_2-\varepsilon_3} \right)\\     
        &+\mathrm{i}2\Delta\left( -\frac{\eta_{41}\phi_{1,12}+\eta_{42}\phi_{2,12}}{\varepsilon_1-\varepsilon_2} 
        \right.\\ &\left.-\frac{\eta_{30}\phi_{0,23}+\eta_{63}\chi_{3,23}}{\varepsilon_2-\varepsilon_3} \right)   ,\\
        \psi_{33}=&2\Delta  \left( \frac{\eta_{20}\phi_{0,13}-\eta_{53}\chi_{3,13}}{\varepsilon_3-\varepsilon_1} 
        +\frac{\eta_{63}\phi_{3,23}}{\varepsilon_2-\varepsilon_3} \right)\\     
        &+\mathrm{i}2\Delta\left( \frac{\eta_{30}\phi_{0,23}+\eta_{63}\chi_{3,23}}{\varepsilon_2-\varepsilon_3} 
        +\frac{\eta_{53}\phi_{3,13}}{\varepsilon_3-\varepsilon_1} \right), 
        \\
    \end{split}
\end{align} 
as also discussed in Eqs.~(\ref{psi_1})-(\ref{psi_2}).
We note that all components of $\hat{\mathcal{D}}_{ij}$ are proportional to $\Delta$.
For later convenience, we additionally apply a unitary transformation:
\begin{align}
\check{\mathcal{H}}_s=
    \check{u}^{\dag}_s
    \check{\mathcal{H}}_r
    \check{u}_s=
   \left(\begin{array}{ccc} \breve{\mathcal{H}}_{\alpha}& 
   \breve{\mathcal{V}}_{\alpha \beta} & \breve{\mathcal{V}}_{\alpha \gamma}  \\ 
\breve{\mathcal{V}}_{\beta \alpha } & \breve{\mathcal{H}}_{\beta} & \breve{\mathcal{V}}_{\beta \gamma}  \\ 
\breve{\mathcal{V}}_{\gamma \alpha} &  \breve{\mathcal{V}}_{\gamma \beta} & \breve{\mathcal{H}}_{\gamma} \end{array} \right), 
\end{align}
with
\begin{align}
\check{u}_s=   
\left(
    \begin{array}{cccccc}
        \tilde{\sigma}_0& 0 & 0 & 0 & 0& 0
        \\
        0 & 0 & 0& 0 & 0& \tilde{\sigma}_0
        \\
        0 & 0 & \tilde{\sigma}_0& 0 & 0& 0
        \\
        0 & \tilde{\sigma}_0 & 0& 0 & 0& 0
        \\
        0 & 0 & 0& 0 & 0& \tilde{\sigma}_0
        \\
        0 & 0 & 0& \tilde{\sigma}_0 & 0& 0
      \end{array}
    \right),     
\end{align}
where
\begin{align}
    \breve{\mathcal{H}}_{\nu}=
        \left(
     \begin{array}{cc}
       \varepsilon_{{\nu}} \tilde{\sigma}_0 & \psi_{{\nu}{\nu}}\mathrm{i}\tilde{\sigma}_y  \\
       -\psi_{{\nu}{\nu}}^*\mathrm{i}\tilde{\sigma}_y & 
       -\varepsilon_{{\nu}} \tilde{\sigma}_0 
     \end{array}
   \right),
\end{align}
\begin{align}
    \breve{\mathcal{V}}_{{\nu}{\nu}'}=
        \left(
     \begin{array}{cc}
       0 & \hat{\mathcal{D}}_{{\nu}{\nu}'}  \\
       -\hat{\mathcal{D}}^*_{{\nu}{\nu}'} & 0
     \end{array}
   \right).
\end{align}
The notations $\nu$ and $\nu'$ represent $\alpha$, $\beta$, or $\gamma$.
The unitary operator $\check{u}$ in Eq.~(\ref{eq:band_ham}) of the main text is defined by
\begin{align}
\check{u}=\check{u}_p\check{u}_q\check{u}_r\check{u}_s.    
\end{align}

Next, we construct a $4\times4$ low-energy effective Hamiltonian for each band by following Ref.~\cite{P-M-R-Brydon,J-W-F-Venderbos}. The Green’s function corresponding to $\check{\mathcal{H}}_s$ is defined by
\begin{align}
    \check{G}_s=(\omega\check{I}-\check{\mathcal{H}}_s)^{-1}
\end{align}
where the Green’s function of the $(\nu,\nu)$ component is given by
\begin{align}
&\breve{G}^{-1}_{\nu}(\bm{k},\omega)=\omega\breve{I}-\breve{H}_{\nu}\nonumber\\
&-
\left(
       \breve{\mathcal{V}}_{\nu'\nu}^\dag 
       \breve{\mathcal{V}}_{\nu''\nu}^\dag
\right)
\left(
\begin{array}{cc}
\omega\breve{I}-\breve{\mathcal{H}}_{\nu'} & 0 \nonumber\\
0 & \omega\breve{I}-\breve{\mathcal{H}}_{\nu''}  
\end{array}
\right)^{-1}\\
& \times \left(
\begin{array}{cc}
       \breve{\mathcal{V}}_{\nu'\nu}  \\
       \breve{\mathcal{V}}_{\nu''\nu}
\end{array}
\right)
+\mathcal{O}(\mathcal{V}^4)\nonumber\\
&=\omega\hat{I}-\hat{\mathcal{H}}_{\nu}^{\mathrm{eff}}(\omega)+\mathcal{O}(\mathcal{V}^4),
\label{eq:Green}
\end{align}
where  
\begin{align}
\breve{\mathcal{H}}^{\mathrm{eff}}_{\nu}(\omega)&=\breve{\mathcal{H}}_{\nu}+
\sum_{\nu'\neq\nu}
\breve{\mathcal{V}}_{\nu'\nu}^\dag (\omega\breve{I}-\breve{\mathcal{H}}_{\nu'})^{-1}\breve{\mathcal{V}}_{\nu'\nu}. 
\end{align}
We only focus on the low-energy excitation around the Fermi surface of the $\nu$ band, where $\varepsilon_\nu\approx0$ is satisfied. In addition, we 
 assume that the $\nu’$ and $\nu^{\prime\prime}$ bands lie energetically far from the Fermi surfaces of the $\nu$ band, which enables us to ignore the effect of the pair potential on the dispersion of $\varepsilon^\prime_\nu$ and $\varepsilon_\nu^{\prime\prime}$ around at $\varepsilon_\nu\approx0$. As a result, we can further approximate $\breve{\mathcal{H}}^{\mathrm{eff}}_{\nu}(\omega)$ 
 as,
\begin{align}    \breve{\mathcal{H}}^{\mathrm{eff}}_{\nu}=\breve{\mathcal{H}}_{\nu}
+\sum_{\nu'\neq\nu}\frac{\breve{\mathcal{V}}^{\dag}_{\nu'\nu} \breve{\tau_z} 
        \breve{\mathcal{V}}_{\nu'\nu}}{\varepsilon_\nu-\varepsilon_{\nu'}} 
\end{align}
which represent the effective Hamiltonian we consider~\cite{P-M-R-Brydon,J-W-F-Venderbos}.
 $\breve{\tau}_z=\mathrm{diag}[1,1,-1,-1]$.
 The similar argument for constructing the effective Hamiltonian from the Green’s function is presented in Ref.~\cite{J-W-F-Venderbos}; for instance, see Eq. (40) of Ref.~\cite{P-M-R-Brydon}. Specifically, the low-energy effective Hamiltonian is represented by
\begin{align}
    \begin{split}
    &    \breve{\mathcal{H}}_{\nu}^{\mathrm{eff}}(\bm{k})=
        \left(
     \begin{array}{cc}
       \tilde{h}_{\nu}(\bm{k}) & \tilde{\Delta}_{\nu} (\bm{k}) \\
       \tilde{\Delta}^{\dag}_{\nu} (\bm{k}) & -\tilde{h}^{\mathrm{T}}_{\nu}(\bm{-k})
     \end{array}
         \right),\\
    &     \tilde{h}_{\nu}=(\varepsilon_{\nu}+\gamma_{\nu})\tilde{\sigma}_0+
    \bm{m}_{\nu}\cdot\tilde{\bm{\sigma}},\\
    &     \tilde{\Delta}_{\nu}=\psi_{\nu\nu} \mathrm{i}\tilde{\sigma}_{y},
    \end{split}
\end{align}
where
\begin{align}
\bm{m}_{\nu} =
\sum_{\nu'\neq \nu}\frac{2\mathrm{Re}[\psi_{ij}\bm{d}^*_{\nu'\nu}]-\mathrm{i}[\bm{d}_{\nu'\nu}\times\bm{d}^*_{\nu'\nu}]}{\varepsilon_\nu-\varepsilon_{\nu'}}  
\end{align}
\begin{align}
    \gamma_{\nu}=\sum_{\nu' \neq \nu}
\frac{|\psi_{\nu'\nu}|^2+|\bm{d}_{\nu'\nu}|^2}{\varepsilon_{\nu}-\varepsilon_{\nu'}},
\end{align}
 which is also given as Eqs.~(\ref{eq:sift}) and (\ref{eq:zeeman}) in the main text. We note that the pseudo-Zeeman field $m_{\nu}$ and the correction in the kinetic energy $\gamma_{\nu}$ is proportional to $\Delta^2$.


\setcounter{figure}{0}
\renewcommand{\thefigure}{C\arabic{figure}}
\renewcommand{\theequation}{C\arabic{equation}}
\section{RECURSIVE GREEN'S FUNCTION}
\label{ap:recursive}
We can deform the BdG Hamiltonian in Eq.~(\ref{eq:BdG}) as
\begin{align} 
\check{H}(\bm{k})=&\check{h}_0(\bm{k_{||}})\nonumber\\
&+\check{t}_{1}(\bm{k_{||}})
\exp(-\mathrm{i}\frac{k_zc}{2})
+\check{t}_{2}(\bm{k_{||}})\exp(-\mathrm{i}k_zc)\nonumber\\
&+\check{t}^{\dag}_{1}(\bm{k_{||}})\exp(\mathrm{i}\frac{k_zc}{2})+\check{t}^{\dag}_{2}(\bm{k_{||}})
\exp(\mathrm{i}k_zc),
\end{align}
where $\bm{k}_{||}$ is the wave number $(k_x,k_y)$   
, and $c$ is the lattice constant of the conventional unit cell in the $z$ direction.
Here $\check{h}_0$ and $\check{t}_{1(2)}$ are the intra-layer element and the inter-layer elements between the (next) nearest-neighbor layers of the Hamiltonian, respectively.
When we employ a real space basis along the $z$-direction, the Hamiltonian with $n$ layers along the $z$-direction is represented by
\begin{align}
H_n=\frac{1}{2}\sum_{\boldsymbol{k}_{\parallel}}
\tilde{C}^{\dagger}_{n,\boldsymbol{k}_{\parallel}}\tilde{H}_n({\boldsymbol{k}_{\parallel}})\tilde{C}_{n,\boldsymbol{k}_{\parallel}}
\end{align}
\begin{align} 
 &\tilde{H}_{n}(\bm{k}_{\parallel})=
  \left(
    \begin{array}{ccccc}
      \check{h}_0                  & 
      \check{t}_1               &
      \check{t}_2               &
      0                       &
                            \\
      \check{t}_1 ^{\dag}               &                        
      \check{h}_0                  &    
      \check{t}_1                &
      \check{t}_2                &
                             \\
      \check{t}_2  ^{\dag}    &   
      \check{t}_1  ^{\dag}      &
      \check{h}_0                  &
      \check{t}_1                & 
         \cdots                     \\ 
      0          &                           
      \check{t}_2 ^{\dag}         &   
      \check{t}_1 ^{\dag}        &
      \check{h}_0                  & 
                              \\                            
                      &   
                      &
                  \vdots       &
                        &                                                   
    \end{array}
    \right)\\
  &\tilde{C}_{n,\bm{k}_\parallel}^{\dag}=
(\check{C}^{\dag}_{1,\bm{k}_\parallel},\check{C}^{\dag}_{2,\bm{k}_\parallel},\cdots \check{C}^{\dag}_{n,\bm{k}_\parallel}),
\end{align}
where
\begin{align}
\check{C}^{\mathrm{T}}_{j,\bm{k}_\parallel}=
({c}_{j,yz\uparrow,\bm{k}_\parallel},
c_{j,zx\uparrow,\bm{k}_\parallel},
c_{j,xy\uparrow,\bm{k}_\parallel},\nonumber\\
c_{j,yz\downarrow,\bm{k}_\parallel},
c_{j,zx\downarrow,\bm{k}_\parallel},
c_{j,xy\downarrow,\bm{k}_\parallel},\nonumber\\
c_{j,yz\uparrow,-\bm{k}_\parallel}^\dagger,
c_{j,zx\uparrow,-\bm{k}_\parallel}^\dagger,
c_{j,xy\uparrow,-\bm{k}_\parallel}^\dagger,\nonumber\\
c_{j,yz\downarrow,-\bm{k}_\parallel}^\dagger,
c_{j,zx\downarrow,-\bm{k}_\parallel}^\dagger,
c_{j,xy\downarrow,-\bm{k}_\parallel}^\dagger
)
\end{align}
%
is the creation operator of the quasi particles in the $j$-th layer.  
We treat the $n$-layers system as
the $n'(=n/2)$ layers system by treating 
two adjacent original layers as a single layer.
The Green's function of the $2n'$ layers system is given by
$\tilde{G}^{n'}(z,\bm{k}_\parallel)={[z\tilde{I}-\tilde{H}_{2n'}(\bm{k}_\parallel)]}^{-1}$,
where $\tilde{I}$ is the unit matrix and  
$z=\varepsilon+\mathrm{i}\delta$.
$\varepsilon$ is a real frequency, and $\delta$ is an infinitesimal imaginary part.
We define $\dot{G}^{n'}_{n'}$ as 
the $(n',n')$ component of $\tilde{G}^{n'}$.
$\dot{G}^{n'}_{n'}$ is a $24\times24$ matrix with the orbital-, spin-, particle-hole-, and layer-degrees of freedom.
By taking the limit $n'\rightarrow\infty$ with the M\"obius transformation, we obtain the top-surface Green's function for the semi-infinite system \cite{A-Umerski}:
$\dot{g}^s=\lim_{{n'}\rightarrow\infty}\dot{G}_{n'}^{n'}$
 which satisfies  
\begin{align}
    \label{eq:converge} 
&\dot{g}^s(\bm{k}_{||},z)=
[
 z\dot{I}-\dot{h}(\bm{k}_{||})
  -\dot{t}^{\dag}(\bm{k}_{||})\dot{g}^s(\bm{k}_{||},z)
   \dot{t}(\bm{k}_{||})
]^{-1},\nonumber\\
&\dot{h} =  
\left(
    \begin{array}{cc}
      \check{h}_0              &
      \check{t}_1                   \\
      \check{t}_1^{\dag}                                          
                            &
      \check{h}_0                      
    \end{array}
  \right)
,\quad \dot{t} =  
\left(
    \begin{array}{cc}
      \check{t}_2              &
      0                   \\
      \check{t}_1                                          
           &
      \check{t}_2                      
    \end{array}
  \right).
\end{align}
In the main text, $h$ and $t$ in Eq.~(\ref{eq:recursive}) correspond to $\dot{h}$ and $\dot{t}$, respectively.

\setcounter{figure}{0}
\renewcommand{\thefigure}{D\arabic{figure}}
\renewcommand{\theequation}{D\arabic{equation}}
\section{(001) SURFACE STATE OF CHIRAL $d$-WAVE SC STATE}
\label{ap:chiral_d}
\begin{figure}[!htb]
 \includegraphics[width=1\linewidth]{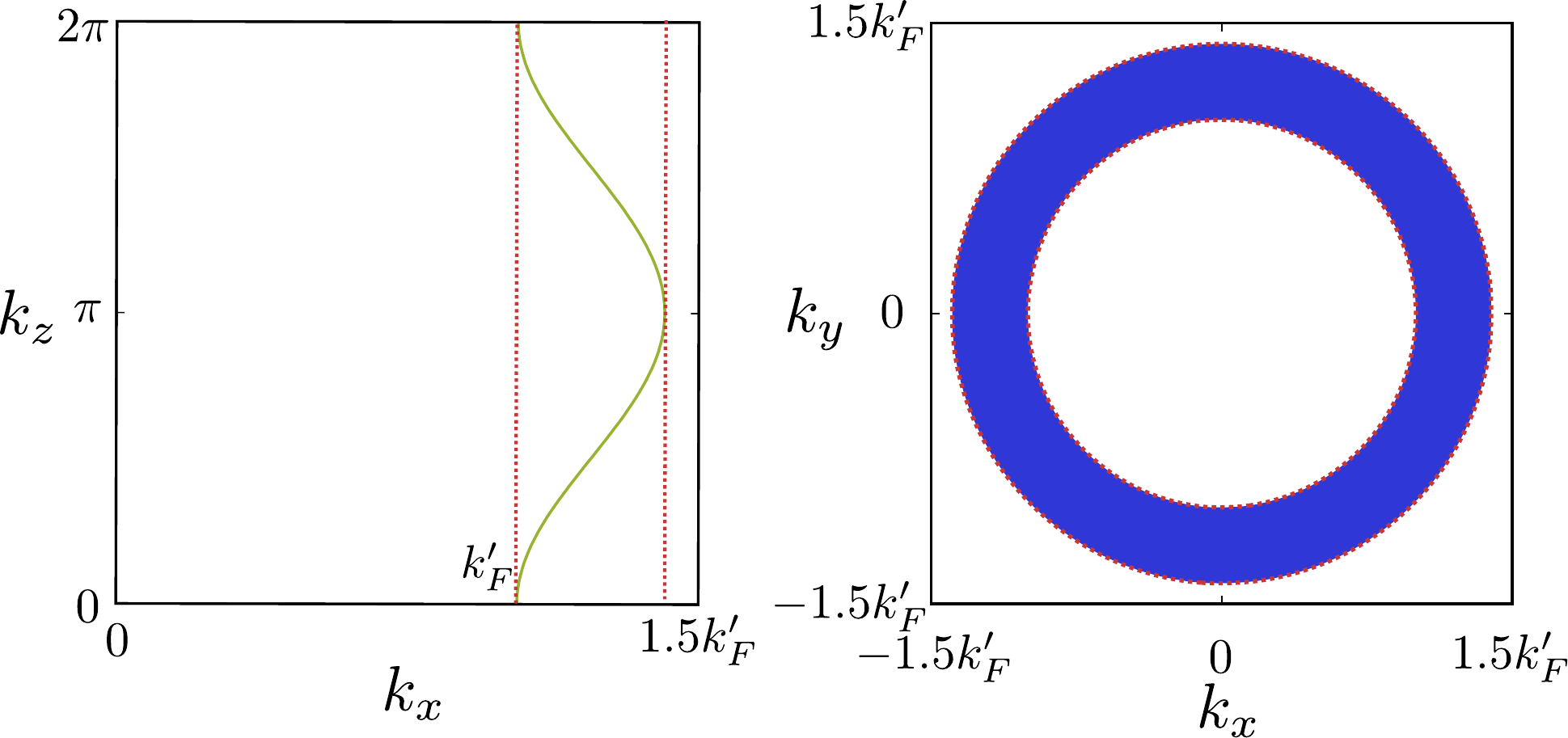}
  \caption{(Color online)
  (a) The $k_z$ dependence of the Fermi surface of Eq.~(\ref{eq:syl_dispersion}) $(t_z=0.1\mu)$ on the $k_x$-axis.
  (b) The momentum range where the zero-energy surface states appear. }
  \label{fig:FS_syl}
\end{figure}
We discuss the (001) surface states of the chiral $d$-wave SC state 
in the single-band model 
by following Ref.~\cite{S-Kobayashi}.
We consider the BdG Hamiltonian,
$\mathcal{H}=\frac{1}{2}\sum_{\bm{k}}\Psi_{\bm{k}}^{\dag}H(\bm{k})
\Psi_{\bm{k}}$ with 
$\Psi^{\mathrm{T}}_{\bm{k}}=
(c_{\bm{k}\uparrow},
c_{\bm{k}\downarrow},
c^\dag_{-\bm{k}\uparrow},
c^\dag_{-\bm{k}\downarrow})$, 
\begin{align}
H(\bm{k})=
\begin{pmatrix}
\epsilon(\bm{k}) & \Delta(\bm{k})
\\
\Delta^\dagger(\bm{k}) &-\epsilon^{\mathrm{T}}(-\bm{k})
\end{pmatrix},
\end{align}
where 
$c_{\bm{k}s}$ is the annihilation operator for a momentum $\bm{k}$ and a spin $s$.
We consider the kinetic energy of the normal state as
\begin{align}
\label{eq:syl_dispersion}
    \epsilon(\bm{k})=\left[\frac{\hbar^2}{2m}(k_x^2+k_y^2)-2t_z(1-\cos{k_z})-\mu\right]\sigma_0.
\end{align}
where $t_z$ is the nearest-neighbor hopping for the $c$-axis, $m$ is the effective mass of electron in the $xy$-plane, $\mu$ is the chemical potential. 
By employing a tight-binding approximation along $z$-direction, we reproduce a cylindrical Fermi surface as shown in Fig.~\ref{fig:FS_syl}(a), where we choose $t_z=0.1\mu$. 
The pair potential is given as
\begin{align}
\Delta(\bm{k})=
\frac{\Delta_0}{k'_F}\sin(k_z)
(k_x +\mathrm{i}k_y) \mathrm{i}{\sigma}_2\nonumber\\
\frac{{\Delta}_0}{k'_F} \sin{k_z} \sqrt{(k_x^2+k_y^2)} e^{\mathrm{i}\phi_{\bm{k}}}\mathrm{i} {\sigma}_2    
\end{align}
where $k'_F=\sqrt{\frac{2m\mu}{\hbar^2}}$ and $\phi_{\bm{k}}=\tan^{-1}(\frac{k_y}{k_x})$. 
The pair potential has sinusoidal dependence with respect to $k_z$ and therefore has the line nodes at $k_z=0$ and $\pi$.
The Hamiltonian has particle hole symmetry: $CH(\bm{k})C^{-1}=-H(-\bm{k})$ with $C={\sigma}_0\tau_1 K$ \cite{S-Kobayashi}.
Here $K$ is a complex conjugation operator and $\tau_{1,2,3}$ are Pauli matrices in the particle-hole space.
By utilizing a local gauge transformation,
$U_{\phi_{\bm{k}}}=e^{-\mathrm{i}\frac{1} {2}\phi_{\bm{k}}\sigma_{0}\tau_3}$,
we can see that the Hamiltonian has pseudo-time reversal symmetry: $U^\dagger_{\phi_{\bm{k}}}TU_{\phi_{\bm{k}}}H(\bm{k})U^\dagger_{\phi_{\bm{k}}}TU_{\phi_{\bm{k}}}=H(-\bm{k})$ with time reversal operator $T=\mathrm{i}{\sigma}_2\tau_0 K$. 
Then, the Hamiltonian has chiral symmetry that the Hamiltonian 
anticommutes with a chiral operator $\Gamma_{\phi_{\bm{k}}}=-\mathrm{i}C U_{\phi_{\bm{k}}}^\dagger T U_{\phi_{\bm{k}}}$: $\{H(\bm{k}),\Gamma_{\phi_{\bm{k}}}\}=0$.
As a result, we can define a one-dimensional winding number $w$ which characterizes the topologically protected surface states at the (001) surface:
\begin{align} \label{eq:wn1}
w(\bm{k}_{\parallel})=\frac{\mathrm{i}}{4\pi}\int d k_z \mathrm{tr}[\Gamma_{\phi_{\bm{k}}}H^{-1}(\bm{k})\partial_{k_{\perp}}H(\bm{k})].
\end{align} 
According to the bulk-boundary correspondence, the nonzero winding number at $\boldsymbol{k}_{\parallel}$ guarantees the zero-energy surface states at $\boldsymbol{k}_{\parallel}$. Thus, when the winding number becomes nonzero in a finite range with respect to $\boldsymbol{k}_{\parallel}$, we obtain the flat-band 
zero-energy surface states at the (001) surface~\cite{M-Sato}. 
 The Equation~(\ref{eq:wn1}) can be rewritten as~\cite{M-Sato}
\begin{align}
\label{eq:winding}
    w(\bm{k}_{||})&=-\frac{1}{2}\sum_{\Delta(k_z\bm{k}_{||})=0}
    \mathrm{sgn}[\partial_{k_z}\Delta(k_z\bm{k}_{||})]\cdot
    \mathrm{sgn}[\epsilon({k_z\bm{k}_{||}})]\nonumber\\
    &=-\frac{1}{2}[\mathrm{sgn}[\epsilon({0,\bm{k}_{||}})]-\mathrm{sgn}[\epsilon({\pi,\bm{k}_{||}})]].
\end{align}
We clearly see that the winding number becomes nonzero in the momentum region where the signs of $\epsilon(0, \boldsymbol{k}_{\parallel})$ and $\epsilon(\pi, \boldsymbol{k}_{\parallel})$ are opposite. Namely, we obtain the flat-band zero-energy states in the momentum region enclosed by the nodal lines projected on the surface BZ as shown in Fig.~\ref{fig:FS_syl}(b). In the main text, we expect that the effective chiral $d$-wave SC   originated from the $E_g$ inter-orbital orbital pair potential can also host the surface states as the pure chiral $d$-wave SC.


\setcounter{figure}{0}
\renewcommand{\thefigure}{E\arabic{figure}}
\renewcommand{\theequation}{E\arabic{equation}}
\section{LDOS UNDER 
 $E_u$ SC STATE  AT THE TOP SURFSCE}
\label{ap:E_u}
In this appendix, we show the stability of the zero-energy peak structure in the LDOS against the chiral $E_u$ pair potential at the surface. Specifically, only at the top surface, we replace the $E_g$ inter-orbital pair potential as in Eq.~(\ref{eq:pairpotential}) in the main text with
\begin{align}
\hat{\Delta}_p&=\left(\begin{array}{cc} 0 & \bar{\Delta}_p \\ \bar{\Delta}_p & 0 \end{array} \right)\nonumber\\
\bar{\Delta}_p &= \Delta_p \bar{\lambda}_p\nonumber\\
\bar{\lambda}_p&=\left(\begin{array}{ccc}
\mathrm{i}\sin{k_ya} & 0 & 0  \\  0 & \sin{k_xa} &   \\  0 &  0 & \sin{k_xa}+\mathrm{i}\sin{k_ya} \end{array} \right) 
\end{align}
where $\Delta_p$ is the magnitude of the pair potential. 
\begin{figure}[!htb]
 \includegraphics[width=1\linewidth]{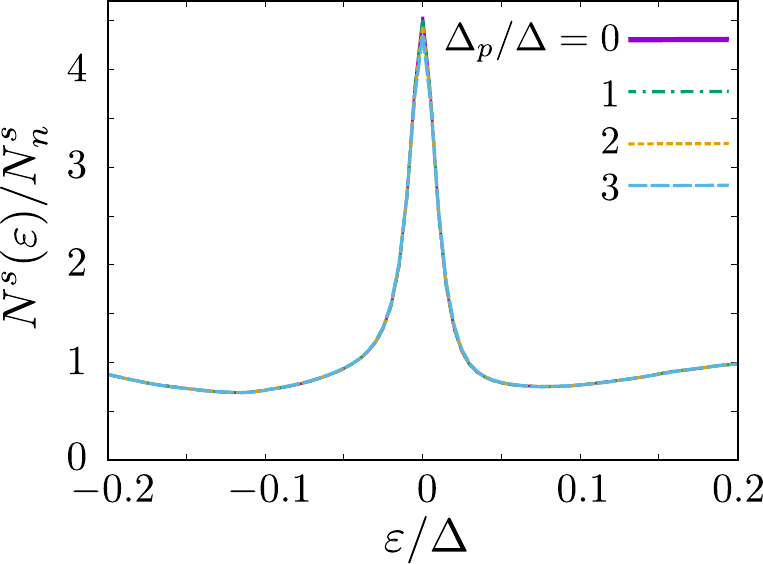}
  \caption{(Color online)
   The $\Delta_p$ dependence of the LDOS is plotted as a function of $\varepsilon$.  There is the chiral $E_u$ pair potenrial 
   state at the top surface instead of the $E_g$ inter-orbital pair potential. $\Delta=|t_z^{(xy,xy)}\times10^{-4}|$.}
  \label{fig:additional_Eu}
\end{figure}
As shown in Fig.~\ref{fig:additional_Eu}, the zero-energy peak structure remains robustly as long as $\Delta_p$ is varied within the order of the bulk pair potential $\Delta$.

\end{document}